\documentclass[12pt]{article}
\usepackage[dvips]{graphicx}
\usepackage[tight]{subfigure}
\usepackage{amssymb}
\usepackage[left=0.1cm,right=2.5cm,top=1.6cm,bottom=1.68cm]{geometry}
\usepackage{blindtext}
\usepackage{geometry}
\def\be{\begin{equation}}
\def\ee{\end{equation}}
\def\bea{\begin{eqnarray}}
\def\eea{\end{eqnarray}}
\def\beaN{\begin{eqnarray*}}
\def\eeaN{\end{eqnarray*}}
\def\ed{\end{document}}
\def\bit{\begin{itemize}}
\def\eit{\end{itemize}}

\def\sig{\sigma}

\def\k{\kappa}
\def\alf{\alpha}

\def\di{\partial}

\def\~{\tilde}
\def\lag{{{\cal L}}}
\def\m{\label}
\def\l{\left}
\def\r{\right}

\def\goto{\rightarrow}

\def\const{\rm const}

\def\sA{{\stackrel{\bullet}{A}}{}}

\def\sR{\stackrel{\bullet}{R}}
\def\sG{\stackrel{\bullet}{\Gamma}}
\def\cG{\stackrel{\circ}{\Gamma}}

\def\sS{{\stackrel{\bullet}{S}}{}}

\def\sK{{\stackrel{\bullet}{K}}{}}
\def\sT{{\stackrel{\bullet}{T}}{}}

\def\sL{\stackrel{\bullet}{\lag}}
\def\cL{\stackrel{\circ}{\lag}}
\def\cN{\stackrel{\circ}{\nabla}}
\def\cA{{\stackrel{\circ}{A}}{}}
\def\cR{\stackrel{\circ}{R}}

\def\scJ{\stackrel{\bullet}{\cal J}}

\def\ccD{\stackrel{}{\cal D}}
\def\ccL{\stackrel{}{\cal L}}
\def\stheta{\stackrel{\bullet}{\theta}}

\usepackage{authblk}

\usepackage[dvipsnames]{xcolor}
\usepackage{graphicx}

\begin{document}

\title{ \bf Mass and angular momentum for the Kerr black hole in TEGR and STEGR}
\author[1,2]{E. D. Emtsova\thanks{Electronic address: \texttt{ed.emcova@physics.msu.ru}}}
\author[3]{A. N. Petrov\thanks{Electronic address: \texttt{alex.petrov55@gmail.com}}}
\author[3,4]{A.V.Toporensky\thanks{Electronic address: \texttt{atopor@rambler.ru}}}

\affil[1]{Physics Department, Ariel University, Ariel 40700, Israel}

\affil[2]{Department of Physics, Bar Ilan University, Ramat Gan 5290002, Israel}
\affil[3]{Sternberg Astronomical institute, MV Lomonosov State university  \protect\\ Universitetskii pr., 13, Moscow, 119992,
Russia}
\affil[4]{Kazan Federal University, Kremlevskaya 18, Kazan, 420008, Russia}
\date{\small \today}
\maketitle

\begin{abstract}
We study the energy-momentum characteristics of the rotating black hole - Kerr solution of general relativity in the Teleparallel Equivalent of General Relativity (TEGR) and the Symmetric Teleparallel Equivalent of General Relativity (STEGR). The previously constructed spacetime covariant and Lorentz invariant expressions for conserved Noether currents, superpotentials and charges are used. The Noether charges describe total energy, momentum or angular momentum of gravitating system depending on a choice of the displacement vector $\xi$. To define covariant and invariant conserved quantities both in TEGR and in STEGR  on needs to use external fields which are flat teleparallel connections. To determine the non-dynamical connections in TEGR and STEGR we use the unified ``turning off'' gravity principle. Besides, to analyse the Noether conserved quantities in these theories, we use the concept of ``gauges''. The gauge changing can affect the Noether conserved quantities. We  highlight two ways to turn off gravity - by $M \to 0$ and by $M \to 0 , ~ a \to 0$ which gives us different gauges in TEGR and STEGR. In both kind of gauges we get the expected values of black hole mass and angular momentum. Our attempts to find gauges which could lead to a correspondence to Einstein's equivalence principle for the Kerr solution where unsuccessful both in TEGR and STEGR. However, these exercises helped us to find a related gauge for the Schwarzschild solution in STEGR that is a novelty.
\end{abstract}

\section{Introduction}
One of the most challenging issues in modern gravitational theory is the compatibility of General Relativity (GR), together with the Standard Model of particle physics, with contemporary cosmological observations. These observations include the accelerated expansion of the Universe at both present times and during its early phases, attributed to dark energy and inflation respectively. Additionally, the observations of galaxies and large-scale structure highlight the presence of an unidentified constituent - dark matter - which manifests through gravitational effects only. Alongside various exotic models of elementary particles, modified theories of gravity offer potential explanations for these phenomena.

Many modifications of GR focus on altering its standard formulation, which is based on Riemannian curvature. However, alternative formulations, such as the Teleparallel Equivalent of General Relativity (TEGR), and the Symmetric Teleparallel Equivalent of General Relativity (STEGR), have also gained significant attention~\cite{BeltranJimenez:2019tjy,Jimenez:2019ghw,Bahamonde:2021gfp}. These modifications are especially intriguing because they have second-order field equations. They also show similarities to gauge field theories, potentially providing connections to theories describing other fundamental interactions in nature. Teleparallel theories of gravity, including TEGR itself \cite{Aldrovandi_Pereira_2013,Maluf,REV_2018}  and its variations like $f(T)$ gravity~\cite{Ferraro:2006jd,Bengochea:2008gz,Ferraro:2008ey,Linder:2010py,Cai:2015emx}, other models~\cite{Hayashi:1979qx,Geng:2011aj,Maluf:2011kf,Bahamonde:2017wwk,Hohmann:2017duq}, and STEGR and it's modifications \cite{Heisenberg:2023lru}, have gained prominence in recent years.

One of important task in any theory is a construction of conservation laws and conserved quantities. Significant efforts in this directions were applied beginning from the pioneering Moller's work   \cite{Moller1961}, see \cite{M_2M,Maluf0704,9,Maluf,REV_2018,Maluf:2018coz,Capozziello2018} and numerous references there in. However, a generally acceptable mechanism for construction of conserved quantities even in the simplest variants of theory, like TERG and STEGR, was not suggested. The corresponding problems are discussed in the book \cite{Aldrovandi_Pereira_2013}. Martin Kr\v s\v s\'ak, in his presentation \cite{Krssak}, summating results in previous studies of many authors, accents on the following. Among important requirements for constructing energy-momentum in TEGR there are two: 1) covariance with respect to both coordinate transformations and local Lorentz rotations, and 2) a possibility to construct well defined global (integral) conserved quantities, or conserved charges. He asserts that in the known approaches it is impossible to satisfy them simultaneously. Note, that many of attention was paid to a problem of a construction of a fully covariant theory, see, for example, a famous review \cite{REV_2018}. However, the above problem was not resolved.

In earlier  works, for example, in TEGR, conserved quantities are obtained by direct reformatting the field equations for deriving the currents and superpotentials and consequent integration that yield the total energy–momentums, or charges. Namely such an approach does not give a simultaneous satisfying the Kr\v s\v s\'ak's requirements. Such a method explore a specific choice of observers which are characterized with the tetrad vectors in the field equations. Of course, this restricts a full covariance of all the conserved quantities including charges.

Such a situation is improved if one applies the Noether procedure with saving the displacement vector $\xi$ with a further its interpretation as a proper vector of an observer, Killing vector of a spacetime, etc. Firstly such an approach has been suggested in the works \cite{Obukhov+,Obukhov_2006,Obukhov_Rubilar_Pereira_2006} where the Kr\v s\v s\'ak requirements are satisfied simultaneously. However, the authors used the  formalism of differential forms that, unfortunately, had not obtain a development.

In our series of works \cite{EPT19,EPT_2020,EP:2021snt,EP:2022ohe,EKPT_2021,EKPT_2021a,EPT:2022uij,EPT:2023hbc}, restricting ourselves to TEGR and STEGR, we have developed a method for constructing conserved quantities applying the Noether theorem as well. However, unlike \cite{Obukhov+,Obukhov_2006,Obukhov_Rubilar_Pereira_2006}, we are working in the framework of the tensorial formalism. Our quantities are fully covariant - coordinate covariant in STEGR and coordinate and Lorentz covariant in TEGR. These quantities gives us fully invariant Noether charges which can represent total energy, momentum or angular momentum. Thus, the Kr\v s\v s\'ak requirements are satisfied simultaneously. Here, in the Introduction, we do not give a description of our formalism. We believe that the detailed presentation given by the Section 2 is preferable. This provides the necessary basis for presenting specific calculations.

Both TEGR and STEGR are dynamically equivalent to GR that means that their field equations are equivalent, thus do not contain teleparallel connections. Therefore, the teleparallel connections do not affect the dynamics of the system at all. They can be fully arbitrary since all their definitions automatically satisfy the related non-tetrad (or non-metric) equations. Nevertherless, teleparallel connections could be restricted by appropriate values of conserved quantities on the level of their construction.

It  should be noted that the situation in modified teleparallel gravity, like  in $f(T)$ and $f(Q)$, is completely different. It is known, for example, that for a spherically symmetric vacuum solution in $f(T)$ only a particular tetrad (so called a “proper tetrad”) is compatible with zero spin connection, so that after fixing a tetrad the allowed spin connection is fixed as well.
So that, if we consider TEGR as a low energy limit of $f(T)$ theory this rises a question about the nature of transition from $f(T)$ to TEGR. Up to present, this is far from being well understood, especially the number of extra degrees of freedom in modified teleparallel gravity in comparison with those in TEGR is still under debate \cite{Nester_2020,Golovnev_2024}. The situation in symmetric teleparallel gravity is even less clear \cite{Ambrosio_2023}. That is a one of reasons why we restrict ourselves by TEGR and STEGR only where a systematization in constructing conserved quantities is required anyway.

Classical solutions play a crucial role as applications of conservation laws in gravitational theory. Our formalism was tested at the set of solutions. Thus, it was obtained the acceptable mass for the Schwarzschild black hole \cite{EPT19,EPT_2020,EP:2021snt,EP:2022ohe,EKPT_2021,EKPT_2021a,EPT:2022uij}. The derivation of the Einstein equivalence principle is the one of important goals at the development of formalism of teleparallel theories \cite{Capozziello:2022zzh}. It has been tested on the backgrounds of the Schwarzschild solution \cite{EKPT_2021,EKPT_2021a,EPT:2022uij}, Friedman-Lemaitre-Robertson-Walker universe \cite{EPT19,EPT_2020,EPT:2022uij},  (Anti-)de Sitter space \cite{EPT19,EPT_2020} and gravitational wave solution \cite{EPT:2023hbc}.

The Kerr solution is the one of the most important models. Upon delving deeper into the literature, it was surprising to discover that there has been relatively little investigation into conserved quantities associated with the Kerr solution \cite{Obukhov+,Xu:2006ki,daRocha-Neto:2002cvu,Maluf:2002zc,Maluf:2000cc,Maluf:1996kx,Maluf:1995rd,Nashed:2006yw}. Our predecessors had primarily focused on TEGR\footnote{The Kerr solution has been found in the framework of the theory $f(T)$ as well \cite{Bejarano2015}, although without considering conserved quantities.}, neglecting its symmetric counterpart, STEGR, entirely in this context. Furthermore, while achieving correct mass calculations for the Kerr solution, attempts to determine its angular momentum yielded inconclusive results.
 Many authors pay their attention to the problem of correspondence to the equivalence principle  \cite{Capozziello:2022zzh,Gomes:2022vrc,BeltranJimenez:2019tme,Rodrigues:2013ifa}, however, for the best of our knowledge, there are no articles where it has being considered in the framework of the Kerr solution.

 All the above stimulates us, both in TEGR and in STEGR, to apply our full covariant formalism for constructing conserved quantities to obtain mass and angular momentum for the Kerr black hole. Another task is to test this formalism for the problem of correspondence to the equivalence principle on the example of the Kerr solution as well. {\em All of these are the main goals of the present paper.} These insights, on the example of the Kerr solution, underscore the importance of our work in expanding the understanding of conserved quantities, although in TEGR and STEGR, addressing critical gaps in the literature and advancing the field significantly.

The paper is organized as follows:

In section 2, we give brief introduction into TEGR and STEGR, introduce the definitions of covariant Noether currents, superpotentials and charges for these theories and describe the “turning off” gravity principle by which we define  the teleparallel connections.

In section 3, we calculate total Noether charges (mass and angular momentum) for the Kerr black hole in TEGR and STEGR. To obtain the teleparallel connections in this solution we use the ``turning off'' gravity principle. We explain the different results of connections obtained by different ways of ``turning off'' gravity in the terminology of ``gauges''. To obtain the Noether mass and angular momentum we use time-like and rotating space-like Killing vectors of geometry.

In section 4, we study the problem of describing the equivalence principle. To enter into the problem deeper we are based on a more appropriate Doran coordinates \cite{Doran:1999gb} where a proper time of freely falling observer coincides with the coordinate time.

In section 5, we discuss the results.

In Appendix, we give very cumbersome and prolonged expressions.

\section{On covariant conserved quantities in teleparallel \\equivalents of GR}
\subsection{TEGR}
\m{ElementsT}

Here, we give the main notions and notations in TEGR, the general form of conservation laws and conserved quantities constructed by us in \cite{EPT19,EPT_2020} and the formulated by us principle of ``switching off'' gravity.

\subsubsection{The theory}

In TEGR, we are based on the Lagrangian of the form in \cite{Aldrovandi_Pereira_2013}:
\be
\sL =   \frac{h}{{2}\kappa}\sT\,
\equiv
\frac{h}{2\kappa} \l(\frac{1}{4} {\sT}{}^\rho{}_{\mu\nu} {\sT}_\rho{}^{\mu\nu} + \frac{1}{2} {\sT}{}^\rho{}_{\mu\nu} {\sT}{}^{\nu\mu}{}_\rho - {\sT}{}^\rho{}_{\mu\rho} {\sT}{}^{\nu\mu}{}_\nu\r)\,,
\m{lag}
\ee
where  in $G=c=1$ units the Einstein constant $\kappa=8 \pi $. The torsion tensor ${\sT}{}^a{}_{\mu\nu}$ is defined as
\begin{equation}\label{tor}
{\sT}{}^a{}_{\mu\nu} = \di_\mu h^a{}_\nu - \di_\nu h^a{}_\mu + {\sA}{}^a{}_{c\mu}h^c{}_\nu - {\sA}{}^a{}_{c\nu}h^c{}_\mu,
\end{equation}
where $h^a{}_\nu$ are the tetrad components (we denote tetrad indexes of a quantity  by Latin letters   and  spacetime indexes by Greek letters),
 connected to metric by
\begin{equation}\label{g_munu}
    g_{\mu\nu}=\eta_{ab} h^a{}_\mu h^b{}_\nu\,,
\end{equation}
thus, $h = \det h^a{}_\nu$.
The inertial spin connection (ISC) ${\sA}{}^a{}_{c\nu}$ makes the torsion (\ref{tor}) be covariant with respect to local Lorentz rotations.


Following to \cite{Aldrovandi_Pereira_2013}, we denote by $\bullet$ all teleparallel quantities, which are constructed using the Weitzenb\"ok connection $\sG{}{}^\alpha {}_{\kappa \lambda}$. By definition
\begin{equation}\label{ISCdef}
    \sA{}^a{}_{b\mu} = -h_b{}^\nu \stackrel{\bullet}{\nabla}_\mu h^a{}_\nu,
\end{equation}
where the covariant derivative $\stackrel{\bullet}{\nabla}_\mu$ is defined by $\sG{}{}^\alpha {}_{\kappa \lambda}$.
The Weitzenb\"ok connection $\sG{}{}^\alpha {}_{\kappa \lambda}$ is flat. This means that the corresponding curvature is equal to zero:
\begin{equation}\label{RiemTEGR}
    \sR{} {}^\alpha {}_{\beta \mu \nu} \equiv \di_\mu \sG{}^\alpha {}_{\beta\nu} - \di_\nu \sG{}^\alpha{}_{\beta \mu} + \sG{}^\alpha{}_{\kappa \mu}\sG{}^\kappa {}_{\beta \nu} - \sG{}^\alpha{}_{\kappa \nu}\sG{}^\kappa {}_{\beta \mu}=0\,.
\end{equation}
We note also that the Weitzenb\"ok connection $\sG{}{}^\alpha {}_{\kappa \lambda}$ is compatible with the physical metric, that is, the corresponding non-metricity is zero:
\begin{equation}
 \stackrel{\bullet}{Q}   {}_{\mu \alf \beta}  \equiv  \stackrel{\bullet}{\nabla}{}_\mu g_{\alf \beta}=0.
\end{equation}

Again, following to \cite{Aldrovandi_Pereira_2013}, quantities denoted by a $\circ$ are constructed with the use of the Levi-Civita connection $\cG{}{}^\alpha {}_{\kappa \lambda}$. Thus,
$\cA^a{}_{b\rho} $ is the usual  Levi-Civita spin connection (L-CSC) defined by
\begin{equation}\label{A}
    \cA{}^a{}_{b\mu} = -h_b{}^\nu \cN_\mu h^a{}_\nu,
\end{equation}
where the covariant derivative $ \cN_\mu$ is constructed with  $\cG{}{}^\alpha {}_{\kappa \lambda}$. It is useful to introduce a contortion tensor defined as:
  \begin{equation}\label{K_A_A}
      \sK^a{}_{b\rho} = \sA^a{}_{b\rho} - \cA^a{}_{b\rho}.
  \end{equation}
At last, recall that the transformation of tetrad indices into spacetime ones and vice versa is performed by contraction with tetrad vectors, for example, $\sK^\rho{}_{\mu\nu}=\sK^a{}_{b\nu}h_a{}^\rho h^b{}_\mu$;  $\cR{}^a{}_{b\mu\mu} = h^a{}_\alf h^\beta{}_b\cR{}^\alf{}_{\beta\mu\nu}$;  etc.
As a result, the contortion tensor is rewritten in the convenient form:
    \begin{equation}\label{tor_K}
    \sK^\rho{}_{\mu\nu}=\frac{1}{2}(\sT_\mu{}^\rho{}_\nu+\sT_\nu{}^\rho{}_\mu -\sT^\rho{}_{\mu\nu}).
\end{equation}

It is instructive to discuss the role of the ISC incorporated into the torsion tensor (\ref{tor}), and, consequently, into the Lagrangian (\ref{lag}). The latter can be rewritten in the other form \cite{Aldrovandi_Pereira_2013}:
\be
{\sL} =   {\cL} -\frac{1}{\kappa}\di_\mu\l(h{\sT}{}^{\nu\mu}{}_\nu\r)\,,
\m{lag+div}
\ee
with the Hilbert Lagrangian
\be
{\cL} =  -\frac{h}{2\kappa}\cR\,.
\m{lag_H}
\ee
Here, $\cR$ is the Riemannian curvature scalar expressed through the tetrad components by (\ref{g_munu}), see \cite{Landau_Lifshitz_1975}. Then, because the TEGR Lagrangian (\ref{lag}) contains ISC in the divergence only, if one varies the action defined by it with respect to tetrad components one obtains the field equations in vacuum
 \be
E_a{}^\rho \equiv \frac{\delta \sL}{\delta h^a{}_\rho} \equiv \frac{\di \sL}{\di h^a{}_\rho} - \di_\sig \l(\frac{\di \sL}{\di h^a{}_{\rho,\sigma}} \r)=0,
 \m{EM+}
 \ee
 where $E_a{}^\rho$ does not contain ISC totally because
 \be
E_a{}^\rho \equiv \frac{\delta \sL}{\delta h^a{}_\rho} \equiv \frac{\delta \cL}{\delta h^a{}_\rho}=0\,.
 \m{EMc}
 \ee
 Thus the TEGR field equations and the GR field equations are equivalent.
 It is important to note the following as well. Varying the action with the Lagrangian (\ref{lag}) with respect to $\sA^a{}_{b\rho} $ (that is included in the divergence only) we obtain $0=0$. This means that ISC cannot be determined in the framework of TEGR itself, it is an external structure. Then, if it is necessary, it can be defined by additional requirements only, for example, by a construction of acceptable conserved quantities for a concrete solution. Returning to discussion of the role of $\sA^a{}_{b\rho} $ on the level of the dynamic equations, one can assert that its presence allows us to present the torsion tensor in a covariant form with respect to local Lorentz rotations, see (\ref{tor}). This form gives a possibility to represent the following tensors in fully and evidently covariant form as well.

 The torsion scalar  in (\ref{lag}) can be rewritten in the form:
\begin{equation}\label{torscalar}
\sT=\frac{1}{2}\,{\sS}_a{}^{\rho\sigma}{\sT}{}^a{}_{\rho\sigma},
\end{equation}
where the teleparallel superpotential ${\sS}_a{}^{\rho\sigma}$  defined as
\begin{equation}\label{super_K}
     {\sS}_a{}^{\rho\sigma}=
     \sK{}^{\rho\sigma} {}_a + h_a{}^{\sigma} \sK{}^{\theta \rho} {}_{\theta} - h_a{}^{\rho} \sK{}^{\theta \sigma} {}_{\theta}\,
\end{equation}
 is an antisymmetric tensor in the last two indexes.

All tensors ${\sT}{}^a{}_{\mu\nu}$, $\sK^{\rho\sigma}{}_{a}$, and ${\sS}_a{}^{\rho\sigma}$, are covariant with respect to both coordinate transformations and local Lorentz transformations.
Note, that working in the covariant formulation of TEGR, the local Lorentz covariance means that the tensorial quantities are transformed covariantly under the simultaneous transformation of both the tetrad and the ISC:
\begin{equation}\label{lroth}
h'^a {}_{\mu} = \Lambda^a {}_b (x) h^b {}_\mu\,,
\end{equation}
\begin{equation}\label{spin_trans}
\sA{}'{}^a {}_{b \mu}=\Lambda {}^a {}_c  (x) \sA{} {}^c {}_{d \mu} \Lambda {}_b {}^d   (x)  + \Lambda {}^a {}_c  (x) \partial_\mu \Lambda {}_b {}^c  (x) ,
\end{equation}
where $\Lambda {}^a {}_c  (x)$ is the matrix of a local Lorentz rotation, and $\Lambda {}_a {}^c  (x)$  is an inverse matrix  of the latter. The operation at the right hand side of (\ref{spin_trans}) tells us that ISC can be equalized to zero by an appropriate local Lorentz transformation. Then, by another local Lorentz rotation it can be represented in the form:
\begin{equation}\label{telcon}
 {\sA}{}^a{}_{c\nu}=\~\Lambda^{a} {}_b \partial_\nu    (\~\Lambda^{-1}){}^b{}_c.
\end{equation}
%

\subsubsection{Noether conserved quantities}
\m{CL_TEGR}

The Noether conserved quantities were derived for TEGR Lagrangian
 (\ref{lag}) in \cite{EPT19,EPT_2020} with applying the Noether theorem. It was used the covariance of the Lagrangian and arbitrary displacement vectors $\xi^\alpha$. The Noether current ${\scJ}{}^{\alf}(\xi) $ is a vector density of the weight +1, and the Noether superpotential ${\scJ}{}^{\alf\beta}(\xi)$ is an antisymmetric tensor density  of the weight +1. For such quantities $\cN_\mu \equiv \partial_\mu$, and, thus, the conservation laws have evidently covariant  form:
  \be
\di_\alf {\scJ}{}^{\alf}(\xi) \equiv \cN_\alf {\scJ}{}^{\alf}(\xi) = 0\,.
\m{CL_current}
\ee
  \be
{\scJ}{}^{\alf}(\xi) = \di_\beta {\scJ}{}^{\alf\beta}(\xi) \equiv \cN_\beta {\scJ}{}^{\alf\beta}(\xi)\,.
\m{CL_c_s}
\ee
The Noether current ${\scJ}{}^{\alf}(\xi)$ has the form
\cite{EPT19,EPT_2020}:
\begin{equation}\label{noethercurrtegr}
    {\scJ}{}^\alf(\xi) =  h(\stheta_\sig{}^\alf+\theta_\sig{}^\alf)
\xi^\sigma + \frac{h}{\k} \sS_{\sig}{}^{\alf\rho}\cN_\rho\xi^\sig\,,
\end{equation}
when the field TEGR equations hold. Here, $\theta_\sig{}^\alf$ is matter energy-momentum and the gravitational Noether energy-momentum tensor $\stheta_\sig{}^\alf$ is
\begin{equation}
     \stheta_\sig{}^\alf  \equiv \frac{1}{\k}\sS_{a}{}^{\alf\rho}{\sK}^a{}_{\sig\rho}- \frac{1}{h}\sL \delta^\alf_\sig \,.
\end{equation}
The related Noether superpotential is
\begin{equation}\label{super}
    {\scJ}{}^{\alf\beta}(\xi) = \frac{h}{\k}{\sS}_a{}^{\alf\beta}h^a{}_\sigma\xi^\sig\,.
\end{equation}

It is important to give a physical interpretation of the quantities presented above. We follow the prescription in \cite{EPT_2020}. First, one has to choose the displacement vector $\xi^\alpha$, it can be Killing vector, proper vector of observer, etc. Second, setting a time coordinate as $t=x^0$ and choosing a space section as $\Sigma := t=\const$,
one can interpret ${\scJ}{}^0(\xi)$ as a density on the section $\Sigma$ of the quantity related to a chosen  $\xi^\alpha$. For example, if $\xi^\alpha$ is a time-like Killing vector it is interpreted as the energy density on $\Sigma$.
The Noether charge can be defined as:
\begin{equation}\label{ICQdiv}
    {\cal P}(\xi) = \int_\Sigma d^3x {\scJ}{}^{0}(\xi)=  \oint_{\di\Sigma} ds_i {\scJ}{}^{0i}(\xi),
\end{equation}
where $\Sigma$ can be finite or infinite.
On the other hand, if $\xi^\alpha$ is an observer's proper vector, then the components ${\scJ}{}^\alpha(\xi)$ can be interpreted as components of the energy-momentum vector measured by such observers.

As one can see, both ${\scJ}{}^{\alf}(\xi)$ and ${\scJ}{}^{\alf\beta}(\xi)$ are explicitly spacetime covariant and Lorentz invariant.
Indeed, the Lorentz invariance of the Noether conserved quantities  and the Lorentz covariance of the superpotential ${\sS}_a{}^{\alf\beta}$ is satisfied only under the simultaneous transformation of the tetrad (\ref{lroth}) and the ISC (\ref{spin_trans}). However, as we see below, one can define different ISCs for one tetrad, what gives us different values of the Noether conserved quantities, \cite{EKPT_2021,EKPT_2021a}. We introduce our own definition of {\em ``gauges'' in TEGR}. A gauge is defined as a set of pairs  (tetrad and ISC) which can be obtained from a given combination of the tetrad and ISC  by an arbitrary covariant Lorentz transformations where tetrad transforms as (\ref{lroth}) and  ISC transforms as (\ref{spin_trans}) simultaneously, and/or arbitrary coordinate transformations. Thus, the gauge in our definition it is the equivalence class of the above related pairs of tetrads and ISCs which of them gives the same conserved quantity.

 Concluding the subsection it is important to remark the following.  It is well known the notion of a so-called Wietzenb\"ock gauge \cite{Aldrovandi_Pereira_2013}. It is the case when zero ISC corresponds to a proper tetrad. We use the word ``gauge'' in different senses: when we say ``Wietzenb\"ock gauge'' we mean the only one  pair  (tetrad and zero ISC) inside the gauge in our definition (the equivalence class). Thus, if we have a concrete gauge in our definition then among of the set of pairs we can find the Wietzenb\"ock pair in any case because the ISC can be cancelled by an appropriate transformation (\ref{spin_trans}) anyway. Concrete values of conserved quantities for all the pairs (including Wietzenb\"ock pair) of this gauge are the same.

\subsubsection{Defining the ISC: ``turning off'' gravity principle}
\m{turning_off_TEGR}

ISCs in TEGR are not dynamical quantities, therefore they are left undetermined \cite{Golovnev:2017dox,BeltranJimenez:2019tjy} from the start. Moreover, their choice is not unique.
To determine the ISC the  ``turning off'' gravity principle was introduced in \cite{EPT19,EPT_2020}. This principle is based on the assumption that Noether's current and superpotential  are proportional to contortion components $\stackrel{\bullet}{K} {}^{a} {}_{c\mu }$ (or, alternatively $\stackrel{\bullet}{T} {}^{\alf} {}_{\mu \nu}$ or $\stackrel{\bullet}{S} {}_{a} {}^{\mu \nu}$) and they  have to vanish in absence of gravity.
Thus, to determine $\stackrel{\bullet}{A} {}^{a} {}_{c\mu }$ in correspondence of this requirement we turn to the formula (\ref{K_A_A}).  It was suggested:

1)  for known GR solution, to choose a convenient tetrad and define $\stackrel{\circ}{A} {}^{a} {}_{c\mu }=-h_b{}^\nu \cN_\mu h^a{}_\nu$;

2) to construct related curvature of Levi-Civita spin connection $\cR{}^i{}_{j\mu\nu}=\di_\mu \cA{}^i{}_{j\nu} - \di_\nu \cA{}^i{}_{j\mu} + \cA{}^i{}_{k\mu}\cA{}^k{}_{j\nu} - \cA{}^i{}_{k\nu}\cA{}^k{}_{j\mu}$;

3) to “switch off” gravity solving  the absent gravity equation $\stackrel{\circ}{R} {}^a {}_{b \gamma \delta}=0$  for parameters of the chosen GR solution;

4) when the parameters satisfying $\stackrel{\circ}{R} {}^a {}_{b \gamma \delta}=0$ are found, we take $\stackrel{\circ}{A} {}^{a} {}_{c\mu }=\stackrel{\bullet}{A} {}^{a} {}_{c\mu }$ for the found parameter values.

\noindent Such a procedure generalizes those of ``turning off'' gravity suggested by other authors.


\subsection{STEGR}
\m{ElementsS}

Here, we give the main notions and notations in STEGR, the general form of conservation laws and conserved quantities constructed by us in \cite{EPT:2022uij,EPT:2023hbc} and the formulated by us principle of ``switching off'' gravity.

\subsubsection{The theory}

Let us turn to the Lagrangian in STEGR in the form given in \cite{BeltranJimenez:2019tjy}:
\begin{equation}\label{STlag}
  \ccL{} = \frac{\sqrt{-g}}{2 \kappa} g^{\mu\nu} (L^{\alpha} {}_{\beta \mu} L^{\beta} {}_{\nu \alpha} - L^{\alpha} {}_{\beta \alpha} L^{\beta} {}_{\mu\nu} ),
\end{equation}
where the disformation tensor $ L^{\alpha} {}_{\mu \nu}$  is defined as
\begin{equation}\label{defLofQ}
    L^{\alpha} {}_{\mu \nu}=\frac{1}{2} Q^{\alpha} {}_{\mu \nu} -\frac{1}{2} Q_{\mu} {}^{\alpha} {}_{\nu}-\frac{1}{2} Q_{\nu} {}^{\alpha} {}_{\mu}\,,
\end{equation}
and the non-metricity tensor $Q_{\alpha \mu \nu}$ is defined as follows:
\begin{equation}\label{defQ}
    Q_{\alpha \mu \nu} \equiv \nabla_\alpha g_{\mu \nu}.
\end{equation}
Here, the covariant derivative $\nabla_\alpha$ is defined with the use of the teleparallel connection $\Gamma^{\alpha} {}_{\mu \nu}$ which is symmetric in lower indexes, denoting it as STC (symmetric teleparallel connection). Thus, the corresponding torsion is zero: $T {}^\alpha {}_{\mu \nu} \equiv \Gamma {}^\alpha {}_{\mu \nu} - \Gamma {}^\alpha {}_{\nu\mu } = 0$, that is STC is torsionless. The STC is flat as well because the curvature tensor is zero:
\begin{equation}\label{defRiem}
    R^{\alpha} {}_{\beta \mu \nu} (\Gamma) = \partial_{\mu} \Gamma^{\alpha} {}_{ \nu \beta} -  \partial_{\nu} \Gamma^{\alpha} {}_{\mu \beta} +  \Gamma^{\alpha} {}_{\mu \lambda}  \Gamma^{\lambda} {}_{\nu \beta} -  \Gamma^{\alpha} {}_{\nu \lambda}  \Gamma^{\lambda} {}_{\mu \beta} =0.
\end{equation}
The decomposition of a STC $\Gamma^{\beta} {}_{\mu \nu}$ into the Levi-Civita connection, contortion and the disformation terms, see \cite{BeltranJimenez:2019tjy}, reduces to
\begin{equation}\label{defL}
    L^{\beta} {}_{\mu \nu} \equiv \Gamma^{\beta} {}_{\mu \nu} - \cG{}^{\beta} {}_{\mu \nu}.
\end{equation}

With the use of (\ref{defLofQ}) - (\ref{defL}) one can rewrite (\ref{STlag}) as
\begin{equation}\label{STlagtoHilbert}
      \ccL{} =   \cL+ \frac{\sqrt{-g} g^{\mu\nu}}{2 \kappa} R_{\mu\nu} +   \ccL{}' .
\end{equation}
The first term is the Hilbert Lagrangian (\ref{lag_H}), however, presented here as a function of metric
\be
{\cL} =  -\frac{\sqrt{-g}}{2\kappa}\cR\,.
\m{lag_H+}
\ee
Let us discuss the role of the second term in (\ref{STlagtoHilbert}). Although it is equal to zero (due to (\ref{defRiem})), if we preserve it in the Lagrangian, we need to vary it. Thus, variation of (\ref{STlagtoHilbert}) together with the second term means the variation of (\ref{STlag}) exactly. But, variation of (\ref{STlag}) with respect to $\Gamma^{\beta} {}_{\mu \nu}$ gives $\Gamma^{\beta} {}_{\mu \nu} = \cG{}^{\beta} {}_{\mu \nu}$. This means that the flat STC $\Gamma^{\beta} {}_{\mu \nu}$ can be equal to the Levi-Civita connection that is not flat in general. Since it is not permissible the second term in (\ref{STlagtoHilbert}) has to be cancelled.
The third term $\ccL'$ is a total divergence:
\begin{equation}\label{defD}
      \ccL'{} =   - \frac{\sqrt{-g}}{2 \kappa}  \cN_\alpha (Q^\alpha-\hat{Q}^\alpha)=   \partial_\alpha \l(- \frac{1}{2 \kappa} \sqrt{-g} (Q^\alpha-\hat{Q}^\alpha)\r)=\partial_\alf  \ccD{}^\alf,
\end{equation}
and $Q_\alpha=g^{\mu\nu} Q_{\alpha \mu \nu}$, $\hat{Q}_\alpha=g^{\mu\nu} Q_{\mu \alpha \nu}$.
Finally, in the STEGR we consider the Lagrangian
\begin{equation}\label{Ls}
    \ccL{} =  - \frac{\sqrt{-g}}{2 \kappa} \cR  + \partial_\alf  \ccD{}^\alf
\end{equation}
  instead of (\ref{STlag}) or (\ref{STlagtoHilbert}), where
\begin{equation}\label{defD2}
    \ccD^{\alpha} \equiv  - \frac{\sqrt{-g}}{2 \kappa}  (Q^\alpha-\hat{Q}^\alpha)\,.
\end{equation}

Because the STEGR Lagrangian contains STC in the divergence only, first, varying the action with the Lagrangian (\ref{Ls}) with respect to the metric components one obtains the GR field equations exactly, stressing the equivalence between STEGR and GR. Second, varying the action with the Lagrangian (\ref{Ls}) with respect to STC $\Gamma^\alf{}_{\mu\nu}$ one obtains $0=0$. That is STC cannot be determined in the framework of STEGR itself. This means that STC, the same as ISC in TEGR, being an external structure, can be defined by additional requirements only, like a construction of conserved quantities with acceptable values.

\subsubsection{Noether conserved quantities}
\m{CL_STEGR}

Analogously to the Noether conserved quantities in TEGR we present conserved quantities in STEGR derived from Lagrangian
 (\ref{Ls}) in \cite{EPT:2022uij} with applying the Noether theorem. The Noether current ${\cal J}{}^{\alf}(\xi) $ and the Noether superpotential ${\cal J}{}^{\alf\beta}(\xi)$ with the same transformation properties as those in TEGR satisfy the conservation laws
  \be
\di_\alf {\cal J}{}^{\alf}(\xi) \equiv \cN_\alf {\cal J}{}^{\alf}(\xi) = 0\,.
\m{CL_current+}
\ee
  \be
{\cal J}{}^{\alf}(\xi) = \di_\beta {\cal J}{}^{\alf\beta}(\xi) \equiv \cN_\beta {\cal J}{}^{\alf\beta}(\xi)\,.
\m{CL_c_s+}
\ee
Analogously to TEGR conserved charge is defined as
\begin{equation}\label{ICQdiv+}
    {\cal P}(\xi) = \int_\Sigma d^3x {\cal J}^{0}(\xi)=  \oint_{\di\Sigma} ds_i {\cal J}^{0i}(\xi)
\end{equation}
with the analogous interpretation.

The total Noether  superpotential of the Lagrangian (\ref{Ls}) in STEGR is derived as
\begin{equation}\label{totalsup}
    {\cal J}{}^{\alpha\beta} = {\cal J}{}_{GR}^{\alpha\beta} + {\cal J}{}_{div}^{\alpha\beta}
\end{equation}
respectively to two parts in the Lagrangian (\ref{Ls}).
The superpotential ${\cal J}{}_{GR}^{\alpha\beta}$ for the Hilbert term is the well known Komar superpotential  \cite{Mitskevich_1969, Petrov_KLT_2017}
\begin{equation}\label{Komarsup}
     {\cal J}{}_{GR}^{\alpha\beta}  = \frac{\sqrt{-g}}{\kappa} \cN{}^{[\alpha} \xi^{\beta]}.
\end{equation}
For the divergent term, see (\ref{defD}) and (\ref{defD2}), the Noether superpotential \cite{EPT:2022uij} is 
\begin{equation}\label{addsupstegr}
 {\cal J}{}_{div}^{\alpha\beta}= \frac{\sqrt{-g}}{\kappa} \delta_{\sigma}^{[\alpha}  (Q^{\beta]}-\hat{Q}^{\beta]}) \xi^\sigma.
\end{equation}

We use concept of {\em gauges in STEGR} that has  been introduced by us in  \cite{EPT:2023hbc}.
We define a pair of coordinates and STC $\Gamma {}^\alpha {}_{\mu\nu}$, with the set of  pairs which are  connected to it by the transformation
\begin{equation}\label{affconntransf}
  \Gamma {}^\alpha {}_{\mu\nu} =\frac{\partial x^\alpha}{\partial \bar{x}{}^{\bar{\alpha}}} \frac{\partial \bar{x}^{\bar{\mu}}}{\partial x^\mu}  \frac{\partial \bar{x}^{\bar{\nu}}}{\partial x^\nu}   \bar{\Gamma}{}^{\bar{\alpha}} {}_{\bar{\mu}\bar{\nu}}  +
  \frac{\partial x^\alf}{\partial \bar{x}{}^{\bar{\lambda}}} \frac{\partial}{\partial x^\mu} \l(\frac{\partial \bar{x}{}^{\bar{\lambda}}}{\partial x^\nu} \r)
\end{equation}
as a {\em ``gauge''}. When we say ``gauge'' we mean the set of all possible coordinates and the values of  STCs in them, such that relation (\ref{affconntransf}) is satisfied under all possible coordinate transformations. Thus, the gauge in STEGR is the equivalence class the same as the definition of gauge in TEGR. In another word, the conserved quantities for the same gauge are covariant or invariant with respect to coordinate transformations. Of course, the Komar superpotential is gauge-independent (\ref{Komarsup}) because it does not contain STC at all.

Concluding this subsection, it is instructive to give the next remark. Usually, in literature, the case of zero STC is called as ``coincident gauge'' \cite{Adak:2011ltj,BeltranJimenez:2022azb} that is related to the unique pair (${x}^{\mu}$,  $0$). Thus, the therm ``coincident gauge'' means the same as ``Wietzenb{\"o}ck gauge'' in TEGR that is the unique case (${x}^{\mu}$,  $0$) is the particular case of the full set of pairs (${x}^{\mu}$,  $\Gamma {}^\alpha {}_{\mu\nu}$) of the gauge as an equivalence class.

\subsubsection{Defining the STC: ``turning off'' gravity principle}
\m{turning_off+}

To define the undetermined STC in STEGR we use the adapted for STEGR  ``turning off" gravity principle \cite{EPT:2022uij}. It is based on the assumption that  $Q_{\alpha \mu \nu}$ and $L^{\alpha} {}_{\mu \nu}$ vanish in the absence of gravity and  $\cR{}^{\alpha} {}_{\beta \mu \nu}$ in GR  vanishes in the absence of gravity too. To find the connection in STEGR there are the following steps:

1) for known GR solution, to construct related Riemann curvature tensor of the Levi-Civita connection:
 $$
   \cR{}^\alpha{}_{\beta\mu\nu}=\di_\mu \cG{}^\alpha{}_{\beta\nu} - \di_\nu \cG{}^\alpha{}_{\beta\mu} + \cG{}^\alpha{}_{\kappa\mu}\cG{}^\kappa{}_{\beta\nu} - \cG{}^\alpha{}_{\kappa\nu}\cG{}^\kappa{}_{\beta\mu};
 $$

2) to ``switch off” gravity solving the absent gravity equation $\stackrel{\circ}{R} {}^\alpha {}_{\beta \mu \nu}=0$ for parameters of the chosen GR solution;

3) when the parameters satisfying $\stackrel{\circ}{R} {}^\alpha {}_{\beta \mu \nu}=0$ are found, we take $\Gamma {}^{\alpha} {}_{ \mu \nu}=\stackrel{\circ}{\Gamma} {}^{\alpha} {}_{\mu \nu}$ for the found parameter values.
\\Torsion of the found connection should be zero automatically because we take it from the Levi-Civita connection for some parameter values, and Levi-Civita connection is always symmetric. Curvature of the found connection should be zero too because we found it from the equation $\stackrel{\circ}{R} {}^\alpha {}_{\beta \gamma \delta}=0$.


\subsection{Appropriate gauges}

Concluding section, we would want to repeat the following. The field equations of both TEGR and STEGR, being dynamically equivalent to the Einstein equations, do not include teleparallel connections totally. Thus, on the level of field equations all ISCs and all STCs automatically satisfy all solutions to TEGR or STEGR and do not play a role in the dynamics of systems at all.
However, a necessity to construct {\em covariant conserved quantities} for this or that solution (model) requires to restrict a possible choice of ISCs and STCs.

Such a situation is not new in metric form of GR, where the construction of conserved quantities meet well known difficulties. The classical conserved quantities are not covariant, like well known energy-momentum complex (Einstein's pseudotensor) and Freud's superpotential, see Chapter 1 in the book \cite{Petrov_KLT_2017}. There are many variants to improve the picture. One of them is the bi-metric representation of GR suggested in \cite{KBL_1997} that gives a possibility, making the use of the Noether theorem, to construct conserved
coordinate covariant current and related superpotential that generalize both Einstein's pseudotensor and Freud's superpotential. The main idea is that an auxiliary background
metric $\bar g_{\mu\nu}$ is introduced saving the Einstein equations in the original form without $\bar g_{\mu\nu}$. However, the current and superpotential depend on $\bar g_{\mu\nu}$ drastically. Moreover the choice of $\bar g_{\mu\nu}$ depends on the solution under consideration. For example, it can be flat background if one considers an isolated system with asymptotically flat metric, cosmological background, if one considers cosmological perturbations, etc. However, even if a concrete system for analyzing is chosen ambiguity in a choice of $\bar g_{\mu\nu}$ can be left.

Return to the covariant conserved quantities in TEGR and STEGR. One can find out that the ``turning off" gravity method in TEGR gives ambiguities  \cite{EKPT_2021,EKPT_2021a} in defining ISCs, and the ``turning off" gravity approach in STEGR gives ambiguities  \cite{EPT:2022uij} in STCs. In TEGR, the result depends on the initial tetrad that we choose, in STEGR, the result depends on the coordinates that we choose.

Conserning TEGR, for each tetrad where we ``turn off" gravity we can obtain different gauge (pairs of tetrad and ISC). In general, these pairs are not connected by (\ref{lroth}) and (\ref{spin_trans}) applied simultaneously. Thus, torsion, contortion and superpotential being expressed in the same tetrad and coordinates, but for different gauges,  are different in each case in general. This gives us different values of conserved quantities. But, sometimes, ``turning off" gravity in different ways we can obtain the same gauge \cite{EP:2021snt,EP:2022ohe}.
Conserning STEGR, disformation and non-metricity being expressed in the same  coordinates can be  different with different STCs and this gives us different values of conserved quantities in  each case.

We stress that, from the one hand, our formalism for constructing conserved quantities in TEGR and STEGR is general one because it presents a general method. On the other hand, it is highly solution-dependent, and thus not generally applicable because gauges have to be determined for each concrete solution separately. Thus, there is no, for example, an unique energy-momentum tensor, like in electrodynamics. But it is not surprising situation in GR by the Einstein equivalence principle.

Finalizing, both in TEGR and in STEGR, one of the main purposes is to find the ISCs or STCs, and, respectively, gauges in which we would have physically meaningful results for the concrete solutions. It is a direct analogy with a search for an appropriate background metric in bi-metric GR. Here, our goal to find appropriate gauges which give physically meaningful conserved quantities for the Kerr black hole.

\section{Kerr black hole mass and angular momentum}
\setcounter{equation}{0}

The Kerr metric in Boyer–Lindquist (BL) coordinates $t$, $r$, $\theta$ and $\phi$  (numerated as 0, 1, 2 and 3) in signature ($-+++$)  has a form:
\begin{equation} \label{metkerr}
ds^2=-\frac{\Delta}{\rho^2} (dt-a \sin^2 \theta d\phi)^2 +\frac{\sin^2 \theta}{\rho^2} [(r^2+a^2)d\phi-a dt]^2 + \frac{\rho^2}{\Delta} dr^2 + \rho^2 d\theta^2,
\end{equation}
where
\begin{equation}
\label{notations1}
    \begin{array}{cccc}
        \Delta= r^2-2Mr+a^2, \\
        \rho^2= r^2+a^2 \cos^2 \theta\,.
    \end{array}
\end{equation}
The components of the time-like Killing vector  in BL coordinates are:
\begin{equation}\label{Killingtime}
\xi^\alpha_t = (-1,~0,~0,~0).
\end{equation}
The components of the rotating space-like Killing vector in BL coordinates are:
\begin{equation}\label{Killingphi}
\xi^\alpha_\phi = (0,~0,~0,~1).
\end{equation}

\subsection{TEGR}

To calculate the Noether charges in TEGR, one needs to refer to formula (\ref{ICQdiv}). Concerning the coordinates in the solution (\ref{metkerr}) one needs in the 01-component of the Noether  superpotential (\ref{super}) only. Thus, the total charge is defined at the infinite radius sphere as
\begin{equation}\label{P_01_tegr}
  {\cal P}(\xi) =  \lim_{r\goto\infty}  \oint_{\di\Sigma} {\scJ}{}^{01}(\xi)d\theta d\phi\,,
\end{equation}
where displacement vectors have to be chosen as (\ref{Killingtime}) and (\ref{Killingphi}) for calculating the total mass and angular momentum. To calculate the TEGR superotentail (\ref{super_K}) included into (\ref{super}) it is necessary to find contorsion (\ref{K_A_A}).
Thus, the first step is to calculate $\cA{}^{{a}} {}_{ b \alf}$ defined in (\ref{A}) for that it is necessary to introduce a tetrad and use the Levi-Civita connection components related to the metric (\ref{metkerr}). The last are written down  Appendix, see (\ref{chrkerr}), since they are useful below but being rather cumbersome.

The simplest tetrad that can be introduced in TEGR is the diagonal tetrad, like one written for the Schwarzschild solution in \cite{EPT19,EKPT_2021}. However it cannot be introduced for the solution (\ref{metkerr}) because it has a cross term. Therefore, for the metric (\ref{metkerr}) we search another more suitable tetrad the form of that was successful  in \cite{EPT:2023hbc}. It is
\begin{equation}\label{tetker1}
  h^A {}_\mu = \left(
\begin{array}{cccc}
 {p_0} & 0 & 0 & {p_4} \\
 0 & {p_1} & 0 & 0 \\
 0 & 0 & {p_2} & 0 \\
 0 & 0 & 0 & {p_3} \\
\end{array}
\right).
\end{equation}
Solving $g_{\mu\nu} = h^A {}_\mu h^B {}_\nu \eta_{AB}$ where $\eta_{AB}$ is a Minkowski metric we get
\bea
 {p_0}&=& \frac{\sqrt{\rho^2-2Mr}}{\rho}, \nonumber\\
  {p_1} &=&\frac{\rho}{\sqrt{\Delta}}, \nonumber\\
  {p_2}& =&\rho, \nonumber\\
  p_3&=& \frac{\sqrt{\Delta} \rho \sin  \theta }{\sqrt{\rho^2-2Mr}}, \nonumber\\
  {p_4}& =& \frac{2 a M r \sin ^2\theta }{\rho \sqrt{\rho^2-2Mr}}.\label{ppp}
\eea

Thus, for the tetrad (\ref{tetker1}) with the Levi-Civita connection (\ref{chrkerr}) we calculate $\cA{}^{{a}} {}_{ b \alf}$ defined in (\ref{A}).  Because it is also cumbersome we derive it  in Appendix as well, it is (\ref{L-CSC1}). It was checked that Riemannian tensor is zero when $M=0$. Thus, noting that
\be
\lim_{M\goto 0}\Delta = (a^2+r^2)
\m{lim}
\ee
in (\ref{L-CSC1}) and directly substituting $M\goto 0$ into  (\ref{L-CSC1}) we got ISC\footnote{It seems that turning off gravity by this way we will have the naked singularity. However, parameter $a$, which takes place in ISC, is just a parameter in the external non-dynamic field with zero curvature, therefore it can take a place.} in TEGR:
\bea
\sA{}^{\hat{2}} {}_{ \hat{1} 1}&=&- \sA{}^{\hat{1}} {}_{ \hat{2} 1} =\frac{a^2\sin\theta\cos\theta}{\rho^2\sqrt{a^2+r^2}}, \nonumber\\
\sA{}^{\hat{2}} {}_{ \hat{1} 2}&=&-  \sA{}^{\hat{1}} {}_{ \hat{2} 2}=\frac{r\sqrt{a^2+r^2}}{\rho^2}, \nonumber\\
 \sA{}^{\hat{3}} {}_{ \hat{1} 3}&=&-  \sA{}^{\hat{1}} {}_{ \hat{3} 3} =\frac{r\sin\theta}{\rho}, \nonumber\\
\sA{}^{\hat{3}} {}_{ \hat{2} 3}&=&-\sA{}^{\hat{2}} {}_{ \hat{3} 3}= \frac{\sqrt{a^2+r^2}\cos\theta}{\rho}\,.
               \label{iscM0}
\eea
We will refer to the  obtained gauge of tetrad (\ref{tetker1}) and ISC (\ref{iscM0}) as a {\em Gauge I}.

To obtain the Noether charges (\ref{P_01_tegr}) in this gauge we calculate teleparallel superpotential $\sS{}_\alpha {}^{01}$ (\ref{super_K}) as main part of the Noether superpotential (\ref{P_01_tegr}). For this we calculate contorsion (\ref{K_A_A}) as a difference of ISC (\ref{iscM0}) and L-CSC (\ref{L-CSC1}) and then (\ref{super_K}).

Assuming that $\sS{}_\alpha {}^{01}$ is contracted with the  timelike Killing vector (\ref{Killingtime}) to obtain the mass, we present the component
\bea
\sS_0{}^{01} &=&\rho ^{-6}\l[ \rho ^{4}\l(r-M\r) - \rho ^{3}r\sqrt{\rho ^{2}-2Mr}- \Delta^{1/2}\rho ^{2}r\sqrt{a^2+r^2} \right. \nonumber \\
&+& \left. r\l(a^2+r^2 \r) \l(\rho ^{2}-2Mr \r)+ \rho ^{2}M\l(a^2-r^2 \r)\r].
\label{1_S010}
\eea
One notes also that for the tetrad (\ref{tetker1}) $ h=\left( r^2+  a^2 \cos ^2\theta \right)\sin \theta $.
Then
the superpotential asymptotic behaviour is as follows
\begin{equation}
 {\scJ} {}^{01} = \frac{M \sin  \theta }{4 \pi }  +O \left(  \frac{1}{r}\right)\,.
 \m{asympt_M1_tegr}
\end{equation}
Integrating   it as in (\ref{P_01_tegr})   we get correct black hole mass:
\begin{equation}\label{P_first_tegr_M}
{\cal P}(\xi_t) =     \lim_{r \goto \infty} \oint {\scJ}{}^{0 1}d \theta d \phi =M.
\end{equation}

Assuming that $\sS{}_\alpha {}^{01}$ is contracted with the  rotating Killing vector (\ref{Killingphi}) to obtain the angular momentum,  we present the component
\be
 \sS_3 {}^{01} = Ma\sin^2\theta\rho ^{-6}\l[\rho^2\l(r^2 - a^2 \r) + 2r^2\l(a^2+r^2 \r)-2\rho^3r^2\l(\sqrt{\rho ^{2}-2Mr}\r)^{-1}  \r].
 \label{1_S013}
 \ee
Then
the superpotential asymptotic behaviour is as follows
\begin{equation}\label{asympt_phi1_tegr}
{\scJ} {}^{01} =  \frac{a M \sin^3\theta }{8 \pi }   +O\left(\frac{1}{r}\right).
\end{equation}
Integrating  it at   the infinite radius sphere  we get uncorrect black hole angular momentum:
\begin{equation}\label{P_first_tegr_aM}
  {\cal P}(\xi_\phi) =     \lim_{r \goto \infty} \oint {\scJ}{}^{0 1}d \theta d \phi = aM/3.
\end{equation}
Thus, the Gauge I is unsuccessful.

Using the same tetrad (\ref{tetker1}) let us try to change the ISC. Another way to turn off gravity is to take both $M \goto 0$ and $a \goto 0$ in (\ref{L-CSC1}). 
Then, we get the Schwarzschild black hole ISC obtained in \cite{EPT19}
\bea
\sA{}^{\hat{2}} {}_{ \hat{1} 2}&=&-  \sA{}^{\hat{1}} {}_{ \hat{2} 2}=1, \nonumber\\
 \sA{}^{\hat{3}} {}_{ \hat{1} 3}&=&-  \sA{}^{\hat{1}} {}_{ \hat{3} 3} =\sin\theta, \nonumber\\
\sA{}^{\hat{3}} {}_{ \hat{2} 3}&=&-\sA{}^{\hat{2}} {}_{ \hat{3} 3}=\cos\theta \,.
               \label{iscsbh}
\eea
The pair of ISC (\ref{iscsbh}) with the tetrad (\ref{tetker1}) gives the {\em Gauge I*}.
Repeating all the steps we obtain the component of the teleparallel superpotential $\sS_0 {}^{01}$ necessary for calculating mass instead of (\ref{1_S010}):
\bea
\sS_0 {}^{01}&=& \rho ^{-6}\l[\rho^2 r\l(a^2 +r^2 +\rho^2 \r) - \rho^4 \l(\sqrt{\rho ^{2}-2Mr}+\Delta^{1/2} \r) \right.\nonumber \\
&-& \left. M\rho^2\l(\rho^2+r^2 - a^2 \r)  - 2Mr^2\l(r^2 + a^2 \r) \r]
\label{1a_S010}
\eea
and component $\sS_3 {}^{01}$ necessary for calculating angular momentum instead of (\ref{1_S013}):
  \be
 \sS_3 {}^{01} = Ma\sin^2\theta\rho ^{-6}\l[\rho^2\l(r^2 - a^2 \r) + 2r^2\l(a^2+r^2 \r)-2\rho^4r\l(\sqrt{\rho ^{2}-2Mr}\r)^{-1}  \r].
 \label{1a_S013}
 \ee
Then, we find that the related asymptotic behaviour of superpotentials related to (\ref{1a_S010}) and (\ref{1a_S013}) coincides with (\ref{asympt_M1_tegr}) and (\ref{asympt_phi1_tegr}), respectively. It is because the asymptotic of ISC (\ref{iscM0}) and ISC (\ref{iscsbh}) coincide at $r\goto \infty$. As a result we obtain in the Gauge I* again the acceptable mass (\ref{P_first_tegr_M}) and unacceptable angular momentum (\ref{P_first_tegr_aM}), the same as for the Gauge I.
Thus, the Gauge I* is unsuccessful as well.


To obtain acceptable result let us try to change the gauge, assuming another tetrad in the form
\begin{equation}\label{tetker2}
    h^A {}_\mu =  \left(
\begin{array}{cccc}
 {q_0} & 0 & 0 & {q_4} \\
 0 & {q_1} & 0 & 0 \\
 0 & 0 & {q_2} & 0 \\
 {q_4} & 0 & 0 & {q_3} \\
\end{array}
\right)
\end{equation}
so that it is more complicated than (\ref{tetker1}), but still is simple enough.
We find its components by solving $g_{\mu\nu} = h^A {}_\mu h^B {}_\nu \eta_{AB}$ again and  express them through  the Kerr metric components only
\bea
 {q_0}&=& \frac{|g_{03}|}{g_{03}}A(g)\l[ g_{00}-B(g) \r], \nonumber\\
  {q_1} &=&\sqrt{g_{11}}, \nonumber\\
   {q_2}& =&\sqrt{g_{22}}, \nonumber\\
     q_3&=& \frac{|g_{03}|}{g_{03}}A(g)\l[ g_{33}+B(g) \r] \nonumber\\
  {q_4}& =& |g_{03}|A(g)   \label{qqq}
\eea
(because the explicit expressions are very cumbersome), where
\bea
 A(g) &=& \sqrt{\frac{g_{33}-g_{00}-2B(g)}{\l(g_{00}+g_{33} \r)^2-4g_{03}^2}}, \nonumber\\
  B(g)& =& \sqrt{g_{03}^2-g_{00}g_{33}}.   \label{AB}
\eea
%

Then we calculated L-CSC for the tetrad (\ref{tetker2}). {It is impossible to include here the evident writing of $\cA{}^{{a}} {}_{ b \alf}$ obtained by the PC calculations in Wolfram Mathematica because it is very bulky. However, turning off gravity by $M \goto 0$ in this complicated L-CSC we get the same ISC (\ref{iscM0}). It is explained simply by the fact that at $M \goto 0$ both the tetrad (\ref{tetker1}) and (\ref{tetker2}) become diagonal and coincide. Thus we obtain the other gauge of the tetrad (\ref{tetker2}) and ISC (\ref{iscM0}). We refer to it as - {\em Gauge II} with the new tetrad.

Another way to turn off gravity is to take both $M \goto 0$ and $a \goto 0$ in the complicated L-CSC. These  results are in  the  ISC (\ref{iscsbh}). Thus we tell that the tetrad (\ref{tetker2}) and ISC (\ref{iscsbh}) presents the {\em Gauge II*}. In the Gauge II and Gauge II* cases we provide the same steps as for the Gauge I and Gauge I* above. Again, it is impossible to include here the evident writing of $\scJ{}^{01}$ since it is very unwieldy. However, we present here the superpotential asymptotic behaviour.
For both the gauges, Gauge II and Gauge II*, they coincide

Thus, one has for the timelike Killing vector (\ref{Killingtime})
\begin{equation}
 {\scJ} {}^{01} = \frac{M \sin  \theta }{4 \pi }  +O \left(  \frac{1}{r}\right)\,.
 \m{asympt_M1_tegr34}
\end{equation}
Integrating   it as in (\ref{P_01_tegr})   we get correct black hole mass:
\begin{equation}\label{P_first_tegr_M34}
{\cal P}(\xi_t) =     \lim_{r \goto \infty} \oint {\scJ}{}^{0 1}d \theta d \phi =M.
\end{equation}
For the rotating Killing vector (\ref{Killingphi}) one has
\begin{equation}\label{asympt_phi1_tegr34}
{\scJ} {}^{01} =  \frac{3a M \sin ^3\theta }{8 \pi }   +O\left(\frac{1}{r}\right).
\end{equation}
Integrating  it at   the infinite radius sphere  we get correct black hole angular momentum:
\begin{equation}\label{P_first_tegr_aM34}
  {\cal P}(\xi_\phi) =     \lim_{r \goto \infty} \oint {\scJ}{}^{0 1}d \theta d \phi = aM.
\end{equation}
Thus, both the Gauge II and Gauge II* give acceptable  mass and angular momentum while Gauge I and Gauge I* give correct mass but fail to reproduce correct angular momentum.


Thus, we obtained an acceptable result for both mass and angular momentum simultaneously, even within two different gauges, which looks amazing. Analogously, in the work \cite{EKPT_2021}, it was shown that the mass of a Schwarzschild black hole was obtained both in the static Schwarzschild gauge and in the e-gauge. Besides, one has to remark that at the Schwarzschild limit when $a\goto 0$ all the gauges I, I*, II and II* go to the static Schwarzschild gauge defined in \cite{EKPT_2021}.

\subsection{ STEGR}

To calculate the Noether charges in STEGR, one needs to refer to formula (\ref{ICQdiv+}). Concerning the coordinates in the solution (\ref{metkerr}) one needs in the 01-component of the total superpotential. Thus, the total charge is defined at the infinite radius sphere as
\begin{equation}\label{ICQdiv++}
  {\cal P}(\xi) =  \lim_{r\goto\infty}  \oint_{\di\Sigma} {\cal J}^{01}(\xi)d\theta d\phi\,,
\end{equation}
where the charge is presented in the form of two parts respectively to the superpotential (\ref{totalsup}). One of them is calculated using the Komar superpotential ${\cal J}^{\alpha\beta}_{GR}$ defined in (\ref{Komarsup}), the other part is calculated using the superpotential ${\cal J}{}^{\alpha\beta}_{div}$ defined in (\ref{addsupstegr}).

The Levi-Civita connection for the metric (\ref{metkerr}) is necessary both for calculating the first part (\ref{Komarsup}) of the general superpotential STEGR, and the second part (\ref{addsupstegr}). Expressions for the coefficients ${\cG}{}^\alpha{}_{\mu\nu}$ have been moved to the Appendix (\ref{chrkerr}).

First, to calculate mass we take the time-like Killing vector defined in (\ref{Killingtime}).
Then, taking the metric (\ref{metkerr}) we get the 01-component of Komar superpotential (\ref{Komarsup}):
\begin{equation}\label{KomartM0}
 {\cal J}^{01}_{GR}=- {\cal J}^{10}_{GR} = -\frac{M\sin\theta}{8\pi}\,\frac{(a^2\cos^2\theta-r^2)(a^2+r^2)}{\rho^4},
\end{equation}
In order to calculate the second part of the total superpotential one needs to define the STC for what it is necesary to turn off gravity. Turning off gravity one first notices that Riemannian tensor becomes zero when $M=0$.
Thus, turning off gravity one has to apply $M=0$ in the Levi-Civita connection calculated with (\ref{metkerr}):

Noting the limit (\ref{lim}) and directly substituting $M\goto 0$ into
(\ref{chrkerr}) one finds STC in STEGR:
\bea
\Gamma{}^{1} {}_{ 1 1} &=& \frac{a^2 r \sin^2\theta}{\rho^2 (a^2+r^2)}, \nonumber\\
\Gamma{}^{1} {}_{ 1 2} &=&
\Gamma{}^{1} {}_{ 2 1} =
\Gamma{}^{2} {}_{ 2 2} = -\frac{a^2 \sin\theta\cos\theta}{\rho^2}, \nonumber\\
\Gamma{}^{1} {}_{ 2 2} &=&-\frac{r (a^2+r^2)}{\rho^2},\nonumber\\
\Gamma{}^{1} {}_{ 33} &=&-\frac{r \sin^2\theta(a^2+r^2)}{\rho^2},\nonumber\\
\Gamma{}^{2} {}_{ 1 1} &=& \frac{a^2  \sin\theta\cos\theta}{\rho^2 (a^2+r^2)}, \nonumber\\
\Gamma{}^{2} {}_{ 1 2} &=&
\Gamma{}^{2} {}_{ 2 1} = \frac{r}{\rho^2}, \nonumber\\
\Gamma{}^{2} {}_{ 3 3} &=&- \frac{(a^2+r^2)\sin\theta\cos\theta}{\rho^2},\nonumber\\
\Gamma{}^{3} {}_{ 1 3} &=&
\Gamma{}^{3} {}_{ 3 1} = \frac{r}{a^2+r^2},\nonumber\\
\Gamma{}^{3} {}_{ 2 3} &=&
\Gamma{}^{3} {}_{ 3 2} = \cot \theta .
\label{connkerrm0}
\eea
The coordinates in (\ref{metkerr}) and the STC (\ref{connkerrm0}) present a definite STEGR gauge, let us call it as {\em Gauge SI}.
The difference (\ref{connkerrm0}) and (\ref{chrkerr}) gives the disformation (\ref{defL}) that is defined through non-merticity as in (\ref{defLofQ}). Another way, one can calculate non-metricity defined in (\ref{defQ}) for the metric (\ref{metkerr}) with (\ref{connkerrm0}).
 As a result,  the 01-component of the second part (\ref{addsupstegr}) of the total superpotential is
\begin{equation}\label{addsupM0}
 {\cal J}^{01}_{div}=- {\cal J}^{10}_{div} = \frac{M\sin\theta}{8\pi}\l[\frac{a^2}{a^2+r^2}+\frac{r^2(\rho^2+\Delta-a^2 -r^2)}{\Delta\rho^2}   \r]\,.
\end{equation}
Combining (\ref{KomartM0}) and (\ref{addsupM0}), one obtains the asymptotic for the 01- component of the total superpotential (\ref{totalsup}) as
\begin{equation}
 {\cal J} {}^{01} = \frac{M \sin  \theta }{4 \pi }  +O \left(  \frac{1}{r}\right)\,.
 \m{asympt_M1}
\end{equation}
It is important to note that both (\ref{KomartM0}) and (\ref{addsupM0}) are included in (\ref{asympt_M1}) in two equal parts ${M \sin  \theta }/{8 \pi }$ and ${M \sin  \theta }/{8 \pi }$.
Integrating   it as in (\ref{ICQdiv++})   we get correct black hole mass:
\begin{equation}\label{intbhmassstegr}
{\cal P}(\xi_t) =     \lim_{r \goto \infty} \oint {\cal J}{}^{0 1}d \theta d \phi =M.
\end{equation}

Then, to calculate the angular momentum we take the space-like Killing vector  defined in (\ref{Killingphi}).
Non-zero components of Komar (\ref{Komarsup}) superpotential are
\begin{equation}\label{KomarphiM0}
 {\cal J}^{01}_{GR}=- {\cal J}^{10}_{GR} = \frac{aM\sin^3\theta}{8\pi}\,\frac{2r^2(a^2+r^2)+\rho^2(r^2-a^2)}{\rho^4}\,.
\end{equation}
The 01-component of the second part (\ref{addsupstegr}) of the total superpotential is absent
\begin{equation}\label{addsupstegrM0phi}
 {\cal J}^{01}_{div}=- {\cal J}^{10}_{div} = 0\,.
\end{equation}
Combining (\ref{KomarphiM0}) and (\ref{addsupstegrM0phi}), one obtains for the asymptotic of the 01- component of the total superpotential
\begin{equation}\label{asympt_phi1}
{\cal J} {}^{01} =  \frac{3 a M \sin ^3\theta }{8 \pi }   +O\left(\frac{1}{r^2}\right).
\end{equation}
Integrating  it at   the infinite radius sphere  we get correct black hole angular momemtum:
\begin{equation}\label{intbhmomstegr}
  {\cal P}(\xi_\phi) =     \lim_{r \goto \infty} \oint {\cal J}{}^{0 1}d \theta d \phi = aM.
\end{equation}

Let us provide another way of turning off gravity.
Let us turn off gravity by $M \goto 0$ and $a \goto 0$ in (\ref{chrkerr}). As a result we get the teleparallel connection of Schwarzschild black hole as in \cite{EPT:2022uij}:
\bea
\Gamma{}^{1} {}_{ 2 2} &=&  -r,\nonumber\\
\Gamma{}^{1} {}_{ 3 3} &=& -r \sin^2\theta,\nonumber\\
\Gamma{}^{2} {}_{ 1 2} &=&
\Gamma{}^{2} {}_{ 2 1} =
\Gamma{}^{3} {}_{ 1 3} =
\Gamma{}^{3} {}_{ 3 1} =\frac{1}{r},\nonumber\\
\Gamma{}^{2} {}_{ 3 3} &=& -\sin  \theta  \cos  \theta,\nonumber\\
\Gamma{}^{3} {}_{ 2 3} &=&\Gamma{}^{3} {}_{ 3 2} =
\cot \theta\,.
\m{GGG}
\eea
 By this way of turning off gravity, we got another gauge with the STC (\ref{GGG}) and the coordinates in (\ref{metkerr}), let us call it as {\em Gauge SI*}.

To calculate black hole mass we take time-like Killing vector (\ref{Killingtime}).
Of course, Komar superpotential is the same (\ref{KomartM0}).
But for the 01-component of the second part (\ref{addsupstegr}) of the total superpotential  with (\ref{Killingtime}) we get
\begin{equation}\label{addsupM1}
 {\cal J}^{01}_{div}=- {\cal J}^{10}_{div} = \frac{\sin\theta\l\{ \rho^2\l[r^2(\Delta-a^2)+ \Delta(\Delta+a^2)-r^4\r]-2\Delta^2r^2+ r^2\rho^4\r\}}{16\pi r\Delta\rho^2}\,.
\end{equation}
Then, for the 01- component of the total superpotential one finds the asymptotic like in (\ref{asympt_M1}) with the equal incorporation from (\ref{KomartM0}) and (\ref{addsupM1}).
Integrating like in (\ref{intbhmassstegr})  at  the infinite radius sphere  we get correct  black hole mass $M$ like in (\ref{intbhmassstegr}).

To calculate the angular momentum we take the rotating  Killing vector (\ref{Killingphi}) again, and
Komar superpotential is the same (\ref{KomarphiM0}).
The second part of the Noether superpotential (\ref{addsupstegr}) with $\xi^\mu_\phi$ vanishes like in (\ref{addsupstegrM0phi}):
\begin{equation}\label{addsupstegrM0phi+}
 {\cal J}^{01}_{div}=- {\cal J}^{10}_{div} = 0\,.
\end{equation}
Then, for the asymptotic of the 01- component of the total superpotential one gets the same result (\ref{asympt_phi1}).
Integrating like in (\ref{intbhmomstegr}) at the infinite radius sphere  we get correct  black hole angular momentum $aM$ like in (\ref{intbhmomstegr}). Thus, two different gauges in STEGR give acceptable result. Remark also that at the Schwarzschild limit when $a\goto 0$ both the gauges SI and SI* go to the gauge for the  Schwarzschild solution in STEGR in \cite{EPT:2023hbc} where the correct mass has been obtained as well.

It is interesting to recall that calculations with the unique Komar superpotential, giving the correct angular momentum itself, give only the half of the expected mass. This fact is known as ``Komar's anomaly'' \cite{Petrov_KLT_2017}. In different approaches this problem is solved in various ways. For example, in \cite{Katz1985} in bi-metric formalism an additional divergence in Lagrangian is introduced. Here, in STEGR the analogous situation takes a place, when the Komar superpotential component (\ref{KomartM0}) is added by the components (\ref{addsupM0}) and (\ref{addsupM1}).

\section{Freely falling observers and the equivalence \\ principle}
\setcounter{equation}{0}
\m{freely_falling}

The way of research is as follows. One has to determine a gauge and define  a proper vector $\xi^\alf$ of freely falling observer. Then, in TEGR, one calculates the superpotential ${\scJ}{}^{\alf\beta}(\xi)$ by (\ref{super}) and, next, by (\ref{CL_c_s}) one obtains the current ${\scJ}{}^{\alf}(\xi)$. In STEGR, one calculates the superpotential ${\cal J}{}^{\alf\beta}(\xi)$ by (\ref{totalsup}) and, next, by (\ref{CL_c_s+}) one obtains the current ${\cal J}{}^{\alf}(\xi)$. The results ${\scJ}{}^{\alf}(\xi)=0$ or ${\cal J}{}^{\alf}(\xi)=0$ mean that appropriate gauges have been found and the correspondence to the equivalence principle is achieved.

\subsection{Freely falling observers in the Boyer-Lindquist based gauges}

From the start we check the above defined gauges based on the BL coordinates.
For stating a correspondence with  the equivalence principle one has to use proper vector of freely falling observer instead of the Killing vectors (\ref{Killingtime}) and (\ref{Killingphi}). Such a vector for the metric (\ref{metkerr}) has components
\be
 \xi^\alf=\left\{-\frac{2 M r \left(a^2+r^2\right)}{\Delta\rho^2}-1,\frac{ \sqrt{2M r}\sqrt{
   a^2+r^2}}{\rho^2},0,-\frac{2a  M r}{\Delta\rho^2}\right\}
   \label{properkerr}
\ee
if observer is falling form the rest at infinity. Our calculations for gauges I, I*, II, II*, SI and SI* give ${\scJ}{}^{\alf}(\xi)\neq 0$ or ${\cal J}{}^{\alf}(\xi)\neq 0$. Thus, a correspondence to equivalence principle fails.

However it is interesting to provide a Schwarzschild limit when $a\goto 0$ for these results. Concerning the gauges I, I*, II and II*, they are going to the unique static Schwarzschild gauge defined in \cite{EKPT_2021}, whereas gauges SI and SI* are going to the STEGR static Schwarzschild gauge defined in \cite{EPT:2022uij}. The components of the proper vector (\ref{properkerr}) are going to the related components in the standard Schwarzschild coordinates.
\be
 \xi^\alf=\left\{\frac{r}{r-2M} ,\sqrt{\frac{2M}{ r}},0, 0\right\}\,.
   \label{properSchw}
\ee
In all the cases it was not found a correspondence to equivalence principle the same as in [22] and [25]. Thus, the results ${\scJ}{}^{\alf}(\xi)\neq 0$ and ${\cal J}{}^{\alf}(\xi)\neq 0$ in the framework of the gauges I, I*, II, II*, SI and SI* are not surprising in the case of the Kerr solution as well.

\subsection{Freely falling observers in the Doran based gauges}

In order to search the gauge related to the equivalence principle it is natural to turn to the presentation of the Kerr solutions in the Doran coordinates \cite{Doran:1999gb}, where the proper time of the observer freely falling from the rest at infinity coincides with the coordinate time. Namely, such a property was taken into account in \cite{EKPT_2021,EKPT_2021a} where for describing of the equivalence principle for the Schwarzschild solution the Lemaitre coordinates have been chosen and related gauge has been introduced. Thus, the Doran metric is
\be
ds^2= - dt^2_d + \l[\frac{\rho}{\sqrt{a^2+r^2_d}}dr_d +\frac{\sqrt{2Mr_d}}{\rho}\l(dt_d -a\sin^2\theta_d d\phi_d \r)\r]^2 +\rho^2d\theta^2_d +(a^2+r^2_d)sin^2\theta_d d\phi_d^2\,.
\label{kerrmetfall}
\ee

One can verify the correspondence of Doran's coordinates to BL coordinates in (\ref{metkerr}):
\bea
 r_d &=& r , \nonumber\\
   \theta_d &=&\theta, \nonumber\\
dt_d &=&dt+ \frac{\sqrt{2Mr} \sqrt{a^2+r^2} }{\Delta} dr, \nonumber\\
d \phi_d &=& d \phi+  \frac{a\sqrt{2Mr}}{\Delta\sqrt{a^2+r^2} }  dr.
 \label{DoranTo BL}
\eea
Because  $r_d =r$ and  $\theta_d =\theta$ the notations (\ref{notations1}) for $\rho$ and $\Delta$ bring their initial sense, and we will use $r$ and  $\theta$ without indexes.

The proper vector of observer freely falling from the rest at infinity (\ref{properkerr})  becomes
\begin{equation}\label{kerrobsfall}
 \xi^\alpha =   \left\{-1, \frac{\sqrt{2Mr}\sqrt{a^2+{r}^2}}{\rho^2},0,0\right\}.
\end{equation}

It is useful to derive the Doran metric (\ref{kerrmetfall}) in the Schwarzschild limit (at $a \goto 0$):
\be
ds^2= -\l(1-\frac{2M}{r} \r) dt^2_d + 2\sqrt{\frac{2M}{r}}dr^2  +r^2 \l( d\theta^2+sin^2\theta d\phi_d^2\r)\,.
\label{PGmetfall}
\ee
It is well known Painliv\'e-Gullstrand (PG) metric \cite{Painlev,Gullstrand} for the Schwarzschild solution. Then the proper vector of observer freely falling from the rest at infinity  becomes
\begin{equation}\label{PGobsfall}
 \xi^\alpha =   \left\{-1, \sqrt{\frac{2M}{r}},0,0\right\}.
\end{equation}

\subsubsection{TEGR}

To enter into the TEGR derivation one needs to introduce the tetrad for the metric (\ref{kerrmetfall}). We follow to the suggestion in \cite{Doran:1999gb}, where such a tetrad is defined as a freely falling tetrad:
\begin{equation}\label{kerrtetfall}
    h^a {}_\mu = \left(
\begin{array}{cccc}
 1 & 0 & 0 & 0 \\
s_0 & s_1 & 0
   & s_4\\
 0 & 0 & s_2 & 0 \\
 0 & 0 & 0 & s_3 \\
\end{array}
\right)\,,
\end{equation}
where the components are
\bea
s_0&=& \frac{\sqrt{2Mr}}{\rho}, \nonumber\\
s_1&=& \frac{\rho}{\sqrt{a^2+{r}^2}}, \nonumber\\
s_2&=& \rho, \nonumber\\
s_3&=& \sqrt{a^2+{r}^2}\sin\theta, \nonumber\\
s_4&=& -\frac{a\sqrt{2Mr}\sin^2\theta}{\rho}.
\label{s_Doran}
\eea

Now,  for tetrad (\ref{kerrtetfall}) and metric (\ref{kerrmetfall})  we obtain  L-CSC by the formula (\ref{A}) and turn off gravity by $M \goto 0$. Then the related ISC acquires the form (\ref{iscM0}) exactly. Thus, we have obtained a new gauge presented by the tetrad (\ref{kerrtetfall}) and ISC (\ref{iscM0}). Call it as {\em Gauge III} for the Kerr solution. Next, by the usual way we calculate (\ref{K_A_A}) and the teleparallel superpotential (\ref{super_K}). We omit here combersome expressions saving the space and derive the components  of ${\sS_\alf{}^{\beta\gamma}}$ in Appendix {\ref{Doran_S}} only.

To calculate current we use formula (\ref{CL_c_s}). Then, the components of the Noether superpotential are necessary for that we use ${\sS_\alf{}^{\beta\gamma}}$ derived in ({\ref{Doran_S}}), the proper vector of freely falling observer (\ref{kerrobsfall}) and determinant of the tetrad (\ref{kerrtetfall}). As a result, we get nonzero components of the Noether superpotential
\bea
{\scJ}{}^{0 1} &=& -{\scJ}{}^{10}=-\frac{ a^2 M \sin ^3\theta}{\rho^2}
,\nonumber\\
{\scJ}{}^{0 3} &=& -{\scJ}{}^{30}= -\frac{a \sqrt{2M r}\sin \theta}{{2}r \sqrt{a^2+r^2} }
,\nonumber\\
{\scJ}{}^{1 3} &=& -{\scJ}{}^{ 31}= \frac{ a M \sin\theta}{\rho^2}\,.
\label{superD}
\eea
Finally, formula (\ref{CL_c_s}) gives the current components for freely falling observer
\begin{equation}
{\scJ}{}^{\alf}=     \left\{\frac{2a^2 M r \sin^3\theta}{\rho^4},0,0,\frac{2a M r \sin\theta}{\rho^4}\right\}\,.
\label{currD}
\end{equation}

Turning off gravity by $M \goto 0$ and  $a \goto 0$ gives another gauge, presented by the tetrad (\ref{kerrtetfall}) and ISC (\ref{iscsbh}). Let us call it as {\em Gauge III*}, where we have significantly longer expressions (therefore we do not present them here) that lead to nonzero current too ( we do not present it here as well). Thus, we did not achieved our goal to describe equivalence principle for the Kerr solution.

Nevertheless, it is useful to consider the above results at the Schwarzschild limit when $a\goto 0$. Then the Doran metric (\ref{kerrmetfall}) goes to the PG metric (\ref{PGmetfall}), the expression for the proper vector (\ref{kerrobsfall})  goes to the proper vector (\ref{PGobsfall}). The tetrad (\ref{kerrtetfall}) becomes
\begin{equation}\label{PGtetfall}
h^a {}_\mu = \left(
\begin{array}{cccc}
 1 & 0 & 0 & 0 \\
 \sqrt{\frac{2M}{r}}  & 1 & 0 & 0 \\
 0 & 0 & r & 0 \\
 0 & 0 & 0 & r \sin\theta \\
\end{array}
\right)\,,
\end{equation}
and for both gauges III and III*  the unique ISC (\ref{iscsbh}) is left only. Thus at the Schwarzschild limit $a\goto 0$ one obtains the unique gauge presented by the tetrad (\ref{PGtetfall}) and ISC (\ref{iscsbh}). Let us call it as {\em Gauge III**}.
The main conclusion is that the current with the proper vector (\ref{PGobsfall}) is equal to zero
\begin{equation}
{\scJ}{}^{\alf}=     \left\{0,0,0,0\right\}\,.
\label{currD0}
\end{equation}
that follows from (\ref{currD}) at $a\goto 0$  expressing a correspondence to the equivalence principle.

Let us compare this result with that given in \cite{EKPT_2021} in the Lemaitre gauge. We apply the transformations from the PG coordinates to the Schwarzschild coordinates
\begin{equation}\label{PGtoS}
dt_d= dt-\frac{2M}{r-2M}dr
\end{equation}
with unchanged other ones. Next we are going to the Lemaitre coordinates ($\tau^*$, $\rho^*$):
\bea
dr &=& -\sqrt{\frac{2M}{r}}d\tau^* + \sqrt{\frac{2M}{r}}d\rho^*,\nonumber\\
dt &=& \frac{r}{r-2M}d\tau^* - \frac{2M}{r-2M}d\rho^*,
\label{StoL}
\eea
where
\begin{equation}\label{rtaurho}
r = r(\tau^*,\rho^*) = \l[\frac{3}{2}\l(\rho^* - \tau^* \r) \r]^{2/3}(2M)^{1/3}\,.
\end{equation}
As a result, one obtains the Lemaitre metric
\begin{equation}\label{Lemmet}
ds^2 = -d\tau^{*2} + \frac{2M}{r(\tau^*,\rho^*)}d\rho^{*2}+r^2(\tau^*,\rho^*)d\theta^2  +r^2(\tau^*,\rho^*)\sin^2\theta d\phi^2\,
\end{equation}
instead of PG metric(\ref{PGmetfall}), and the diagonal tetrad
\begin{equation}\label{Ltetfall}
h^a {}_\mu = \left(
\begin{array}{cccc}
 1 & 0 & 0 & 0 \\
 0   & \sqrt{\frac{2M}{r(\tau^*,\rho^*)}} & 0 & 0 \\
 0 & 0 & r(\tau^*,\rho^*) & 0 \\
 0 & 0 & 0 & r(\tau^*,\rho^*) \sin\theta \\
\end{array}
\right)\,
\end{equation}
instead of the tetrad (\ref{PGtetfall}). The the proper vector (\ref{PGobsfall}) of freely falling observer acquires the form
\begin{equation}\label{rtaurho}
\xi^{\alf}  = \left\{-1, 0,0,0 \right\}\,.
\end{equation}
The ISC (\ref{iscsbh}) is preserved because the transformations (\ref{PGtoS}) and (\ref{StoL}) do not change angular coordinates.

Returning to \cite{EKPT_2021} one reminds that the tetrad (\ref{Ltetfall}) and ISC (\ref{iscsbh}) represent the Lemaitre gauge in TEGR. For it the observer with the proper vector (\ref{rtaurho}) measures zero, thus, there is a correspondence with the equivalence principle. This means that gauge III** represents the Lemaitre gauge and, thus, the result (\ref{currD0}) is not surprising.

\subsubsection{STEGR}

To enter to the STEGR derivation one has to follow the standard way defining the gauges related to the coordinates in (\ref{kerrmetfall}). First, one calculates the Levi-Civita connection for (\ref{kerrmetfall}). Second, one turns off gravity by two ways: 1) $M\goto 0$, and 2) $M\goto 0$ with $a\goto 0$ obtaining two different STCs which exactly coincide with (\ref{connkerrm0}) and (\ref{GGG}), respectively. In the result, one has two different gauges, they are coordinates of (\ref{kerrmetfall}) with STC (\ref{connkerrm0}) and with STC (\ref{GGG}). Call them as {\em Gauge SII} and {\em Gauge SII*}. Then, for both the gauges we choose the proper vector (\ref{kerrobsfall}) and  calculate the superpotential (\ref{totalsup}), and next, by the formula (\ref{CL_c_s+}), we calculate current ${\cal J}^\alf$. For both the gauges, the current turns out to be nonzero. Thus, in STEGR, we did not describe equivalence principle for the Kerr solution as well.

We do not derive the very prolonged expressions, however, note that all the components of currents are proportional to the angular momentum parameter $a$. This means that at the Schwarzschild limit when $a\goto 0$ one has to obtain a zero current
\begin{equation}
{\cal J}{}^{\alf}=     \left\{0,0,0,0\right\}\,,
\label{currPG0}
\end{equation}
and, consequently, a correspondence to equivalence principle.

Let us consider the Schwarzschild limit in detail. The Doran metric (\ref{kerrmetfall}) becomes the PG metric (\ref{PGmetfall}). The related  Levi-Civita connection has nonzero components
\bea \label{L-C-PG}
  \cG{}^{0} {}_{ 0 0} &=&
\frac{\sqrt{2} M^{3/2}}{r^{5/2}},\nonumber\\
\cG{}^{0} {}_{ 0 1}&=&
\cG{}^{0} {}_{ 1 0} =
\frac{M}{r^2},\nonumber\\
\cG{}^{0} {}_{ 1 1} &=&
\sqrt{\frac{M}{2r^3}},\nonumber\\
\cG{}^{0} {}_{ 2 2} &=&- \sqrt{2M r},\nonumber\\
\cG{}^{0} {}_{ 3 3} &=&
- \sqrt{2M r} \sin ^2\theta ,\nonumber\\
\cG{}^{1} {}_{ 0 0} &=&
\frac{M (r-2 M)}{r^3},\nonumber\\
\cG{}^{1} {}_{ 0 1} &=&
\cG{}^{1} {}_{ 1 0} =
-\frac{\sqrt{2} M^{3/2}}{r^{5/2}},\nonumber\\
\cG{}^{1} {}_{ 1 1} &=&
-\frac{M}{r^2},\nonumber\\
\cG{}^{1} {}_{ 2 2} &=& 2 M-r,\nonumber\\
\cG{}^{1} {}_{ 3 3} &=&
(2 M-r) \sin ^2\theta ,\nonumber\\
\cG{}^{2} {}_{ 1 2} &=&
\cG{}^{2} {}_{ 2 1} =
\cG{}^{3} {}_{ 1 3} =
\cG{}^{3} {}_{ 3 1} =
\frac{1}{r},\nonumber\\
\cG{}^{2} {}_{ 3 3} &=&
-\sin \theta  \cos \theta ,\nonumber\\
\cG{}^{3} {}_{ 2 3} &=&
\cG{}^{3} {}_{ 3 2} =
\cot \theta .
\eea
The proper vector (\ref{kerrobsfall}) becomes (\ref{PGobsfall}).  The turning of gravity by $M\goto 0$ gives the unique STC in the form (\ref{GGG}). Thus, the coordinates in (\ref{PGmetfall}) and STC (\ref{GGG}) presents a new gauge in STEGR that unites gauges  SII and SII* at the limit $a\goto 0$. Let us call it as {\em SII**}. Next, for the PG metric (\ref{PGmetfall}) with the Levi-Civita connection (\ref{L-C-PG}) and STC (\ref{GGG}), choosing the proper vector as (\ref{PGobsfall}), one calculates the components of the superpotential (\ref{totalsup}) with the parts (\ref{Komarsup}) and (\ref{addsupstegr}). The nonzero components are
\bea
{\cal J}{}^{0 1} &=& -{\cal J}{}^{10}= -\frac{M \sin \theta }{16 \pi }
\label{superPG}
\eea
Applying the formula (\ref{CL_c_s+}) we obtain the result (\ref{currPG0}) and, thus, stating a correspondence with the equivalence principle for the Schwarzschild solution in STEGR. We consider this as a  new result. For the best of our knowledge, there are no articles where such a correspondence has been stated in various approaches. Our attempts to discover this correspondence in \cite{EPT:2022uij} failed because the procedure of turning gravity applied to the Lemaitre solution (\ref{Lemmet}) turns out to be ill-determined and, consequently, the related gauge is ill-determined too. Here, we found out the well defined  gauge
related to this situation, it is the gauge SII**.

\section{Concluding remarks}

Applying our formalism we have obtained both correct mass and angular momentum both in TEGR and in STEGR for the Kerr solution.
  Although we did not achieve the success in description of the equivalence principle, we present our results keeping in mind that it could be instructive for further studies. Moreover, on the basis of our present results we have discovered the equivalence principle in STEGR for the Schwarzschild solution that was not found before.

 In \cite{Obukhov+,Xu:2006ki,daRocha-Neto:2002cvu,Maluf:2002zc,Maluf:2000cc,Maluf:1996kx,Maluf:1995rd,Nashed:2006yw}, the construction of conserved quantities for the Kerr black hole in TEGR was considered. They construct the conserved quantities by direct integration of field equations, from which they obtain non-covariant energy-momentum tensor of gravity. Integrating this energy-momemtum tensor on the hypersurface of the constant time they obtain total energy-momoentum.  Such quantities are not covariant for local Lorentz rotations. In \cite{Maluf:2002zc,Maluf:2000cc,Maluf:1996kx,Maluf:1995rd} the non-covariant formulation of TEGR is used. In \cite{Obukhov+} the formulation of TEGR is covariant, however, the definition of ISC is different. As a result, in all these works they obtain the correct black hole mass. Consideration of Kerr black hole angular momentum was announced in \cite{Maluf:2002zc}, but a concrete result for the Kerr solution was not obtained. We also did not find any literature with definition of conserved quantities for the Kerr black hole in STEGR.

 In the present work, we obtain both mass and angular momentum for the Kerr black hole in both TEGR and STEGR. Our conserved quantities are fully covariant. Note, that our success in constructing conserved charges is achieved due to that we use directly Killing vectors, unlike other approaches.

 We turn off gravity by two different ways - $M \to 0$ and $M\to 0,~~a\to 0$. In TEGR, we use two simplest tetrads for the Kerr black hole metric. Turning off gravity by two different ways for each tetrad we get four different gauges. In two gauges associated with the first tetrad, with the Killing vector (\ref{Killingtime}) we got the correct black hole mass $M$; however, with the Killing vector (\ref{Killingphi}) we got the non-correct black hole angular momentum  $aM/3$. In two gauges associated with the second tetrad, with the Killing vector (\ref{Killingtime}) we got the correct black hole mass $M$; with the Killing vector (\ref{Killingphi}) we got the  correct black hole angular momentum  $aM$. In STEGR, this leads to two  different gauges. In each gauge, with the Killing vector (\ref{Killingtime}) we got the correct black hole mass $M$; with the Killing vector (\ref{Killingphi}) we got the correct black hole angular momentum  $aM$.

At last, our attempts to describe the equivalence principle for the Kerr solution became unsuccessful. In both TEGR and STEGR the correponding current is not zero for a free falling observer even
if we use Doran coordinate system, which is constructed to describe a free fall and is regular at the horizon. For the Doran tetrad (or connections) the non-zero components of the current
are proportional to the parameter $a$, so it is the angular moment of the black hole which is responsible for this discreapance. It is interesting because while in TEGR $a=0$ limit is very well
described by Schwarzschild soution with corresponding Lemaitre tetrad, in STEGR this limit did not allow us to get a satisfactory result when working directly in the Schwarzschild geometry \cite{EPT:2022uij}.
On the contrary, starting from Kerr solution and making the $a \to 0$ limit gives well defined zero current.

\bigskip

{\bf Acknowledgments.}
EE has been supported in part  by  the ministry of absorption and the ``Program
of Support of High Energy Physics'' Grant by Israeli Council for Higher
Education and EE is supported by the Israel Science Fund (ISF) grant No. 1698/22.
AP and AT has been supported by the Interdisciplinary Scientific and
Educational School of Moscow University “Fundamental and Applied Space Research”. The work of AT  have been supported by  the Russian Government Program of Competitive Growth of Kazan Federal University.

\appendix

\section{Appendix}
\setcounter{equation}{0}
Here, we give components of various fields to save a place in a main body of the paper.

The Levi-Civita connection related to the Kerr metric (\ref{metkerr}) in BL coordinates has the components:
\bea \label{chrkerr}
\cG{}^{0} {}_{ 0 1} &=& \cG{}^{0} {}_{10}=-\frac{M(a^2+r^2)(\rho^2-2r^2)}{\Delta\rho^4},\nonumber\\
\cG{}^{0} {}_{ 0 2} &=& \cG{}^{0} {}_{20}=-\frac{2a^2Mr \sin\theta\cos\theta}{\rho^4},\nonumber\\
\cG{}^{0} {}_{ 1 3} &=&
\cG{}^{0} {}_{ 3 1} = \frac{aM\sin\theta\l[\rho^2(a^2 - r^2) -2r^2(a^2+r^2) \r]}{\rho^4},\nonumber\\
\cG{}^{0} {}_{ 2 3} &=&
\cG{}^{0} {}_{ 3 2} = \frac{2a^3Mr\sin^3\theta\cos\theta}{\rho^4},\nonumber\\
\cG{}^{1} {}_{ 0 0} &=& - \frac{M\Delta(\rho^2- 2 r^2)}{\rho^6},\nonumber\\
\cG{}^{1} {}_{ 0 3} &=&
\cG{}^{1} {}_{ 3 0} =\frac{aM\sin^2\theta\Delta(\rho^2- 2 r^2)}{\rho^4},\nonumber\\
\cG{}^{1} {}_{ 1 1} &=& \frac{a^2r\sin^2\theta + M(\rho^2-2r^2)}{\Delta\rho^2},\nonumber\\
\cG{}^{1} {}_{ 1 2} &=&
\cG{}^{1} {}_{ 2 1} =
\cG{}^{2} {}_{ 2 2} = -\frac{a^2\sin\theta\cos\theta}{\rho^2},\nonumber\\
\cG{}^{1} {}_{ 2 2} &=& -\frac{r\Delta}{\rho^2},\nonumber\\
\cG{}^{1} {}_{ 3 3} &=& -\frac{\Delta\sin^2\theta\l\{\rho^4 r -M\l[(\rho^2 -r^2)^2 - (\rho^2 -r^2)(a^2+r^2) +a^2r^2 \r] \r\}}{\rho^6},\nonumber\\
\cG{}^{2} {}_{ 0 0} &=& -\frac{2a^2Mr\sin\theta\cos\theta}{\rho^6},\nonumber\\
\cG{}^{2} {}_{ 0 3} &=&
\cG{}^{2} {}_{ 3 0} = \frac{2aMr(a^2+r^2)\sin\theta\cos\theta}{\rho^6},\nonumber\\
\cG{}^{2} {}_{ 1 1} &=& \frac{a^2\sin\theta\cos\theta}{\Delta\rho^2},\nonumber\\
\cG{}^{2} {}_{ 1 2} &=&
\cG{}^{2} {}_{ 2 1} = \frac{r}{\rho},\nonumber\\
\cG{}^{2} {}_{ 3 3} &=& -\frac{\sin\theta\cos\theta\l\{\rho^4 (a^2+r^2) -2Mr\l[\rho^4 - (a^2+r^2)^2  \r] \r\}}{\rho^6},\nonumber\\
\cG{}^{3} {}_{ 0 1} &=&
\cG{}^{3} {}_{ 1 0} = -\frac{aM(\rho^2-2r^2)}{\Delta\rho^4},\nonumber\\
\cG{}^{3} {}_{ 0 2} &=&
\cG{}^{3} {}_{ 2 0} = -\frac{2aMr\cot\theta}{\rho^4},\nonumber\\
\cG{}^{3} {}_{ 1 3} &=&
\cG{}^{3} {}_{ 3 1} = \frac{\rho^4 r -M\l[(\rho^2 -r^2)^2 - (\rho^2 -r^2)(r^2- a^2) + r^2(a^2+2r^2) \r] }{\Delta\rho^4},\nonumber\\
\cG{}^{3} {}_{ 2 3} &=&
\cG{}^{3} {}_{ 3 2} = \frac{\cot\theta\l[\rho^4  -2Mr(\rho^2 -a^2-r^2) \r] }{\rho^4}\,.
\eea

The L-CSC components for the tetrad (\ref{tetker1}) with the Levi-Civita connection (\ref{chrkerr}) are
\bea \label{L-CSC1}
    \cA{}^{\hat{0}} {}_{ \hat{1} 0} &=&
\cA{}^{\hat{1}} {}_{ \hat{0} 0} = -\frac{M(\rho^2-2r^2)\Delta^{1/2}}{\rho^4\sqrt{\rho^2-2Mr}},\nonumber\\
\cA{}^{\hat{0}} {}_{ \hat{1} 3} &=&
\cA{}^{\hat{1}} {}_{ \hat{0} 3} =\frac{aM\sin^2\theta(\rho^2-2r^2)\Delta^{1/2}}{\rho^4\sqrt{\rho^2-2Mr}},\nonumber\\
\cA{}^{\hat{0}} {}_{ \hat{2} 0} &=&
\cA{}^{\hat{2}} {}_{ \hat{0} 0} =-\frac{2a^2Mr\sin\theta\cos\theta}{\rho^4\sqrt{\rho^2-2Mr}},\nonumber\\
\cA{}^{\hat{0}} {}_{ \hat{2} 3} &=&
\cA{}^{\hat{2}} {}_{ \hat{0} 3} =\frac{2aMr(a^2+r^2)\sin\theta\cos\theta}{\rho^4\sqrt{\rho^2-2Mr}},\nonumber\\
\cA{}^{\hat{0}} {}_{ \hat{3} 1} &=&
\cA{}^{\hat{3}} {}_{ \hat{0} 1} =-\frac{aM\sin\theta(\rho^2-2r^2)}{\Delta^{1/2}\rho^2{(\rho^2-2Mr)}},\nonumber\\
\cA{}^{\hat{0}} {}_{ \hat{3} 2} &=&
\cA{}^{\hat{3}} {}_{ \hat{0} 2} = \frac{2aM\cos\theta\Delta^{1/2}}{\rho^2{(\rho^2-2Mr)}},\nonumber\\
\cA{}^{\hat{1}} {}_{ \hat{2} 1} &=&
-\cA{}^{\hat{2}} {}_{ \hat{1} 1} = -\frac{a^2\sin\theta\cos\theta}{\Delta^{1/2}\rho^2},\nonumber\\
\cA{}^{\hat{1}} {}_{ \hat{2} 2}& =&
-\cA{}^{\hat{2}} {}_{ \hat{1} 2} =  -\frac{r\Delta^{1/2}}{\rho^2},\nonumber\\
\cA{}^{\hat{1}} {}_{ \hat{3} 0} &=&
-\cA{}^{\hat{3}} {}_{ \hat{1} 0} =
\frac{aM\sin\theta(\rho^2-2r^2)}{\rho^2{(\rho^2-2Mr)}},\nonumber\\
\cA{}^{\hat{1}} {}_{ \hat{3} 3} &=&
-\cA{}^{\hat{3}} {}_{ \hat{1} 3} =
-\frac{\sin\theta\l\{\rho^4 r -M\l[(\rho^2 -r^2)^2 - (\rho^2 -r^2)(r^2- a^2) + r^2(a^2+2r^2) \r]\r\} }{\rho^4\sqrt{\rho^2-2Mr}},\nonumber\\
\cA{}^{\hat{2}} {}_{ \hat{3} 0} &=&
-\cA{}^{\hat{3}} {}_{ \hat{2} 0} =
\frac{2aM\cos\theta\Delta^{1/2}}{\rho^4\sqrt{\rho^2-2Mr}},\nonumber\\
\cA{}^{\hat{2}} {}_{ \hat{3} 3} &=&
-\cA{}^{\hat{3}} {}_{ \hat{2} 3} =
-\frac{\cos\theta\Delta^{1/2}\l[\rho^4  -2Mr(\rho^2 -a^2-r^2) \r] }{\rho^4\sqrt{\rho^2-2Mr}}\,.
\eea

The teleparallel superpotential components in all coordinate indexes (\ref{super_K}) calculated for gauge III:
\bea
\sS{}_{{0}} {}^{0 1} &=& -  \sS{}_{{0}} {}^{10} =\frac{M\l[\rho^2\l(\rho^2-a^2+r^2 \r)+2r^2\l(a^2+r^2 \r)  \r]}{\rho^6}
,\nonumber\\
\sS{}_{{0}} {}^{0 2} &=& - \sS{}_{{0}} {}^{20} =-\frac{2Mra^2\sin\theta\cos\theta}{\rho^6}
,\nonumber\\
\sS{}_{{0}} {}^{0 3} &=& - \sS{}_{{0}} {}^{30} =\frac{a\sqrt{2 Mr}}{r\rho^2\sqrt{a^2+r^2} }
,\nonumber\\
\sS{}_{{0}} {}^{1 3}& =& - \sS{}_{{0}} {}^{31} =
 -\frac{ a M \left(\rho^2- 2r^2\right)}{\rho^6}
 ,\nonumber\\
\sS{}_{{0}} {}^{2 3} &=& - \sS{}_{{0}} {}^{3 2} = -\frac{2 M ra \cot \theta}{\rho^6}
 ,\nonumber\\
 \sS{}_{{1}} {}^{0 1} &=& - \sS{}_{{1}} {}^{1 0}= -\frac{ r \sqrt{2M r} \left(\rho^2 + a^2+ r^2\right)}{\rho^4\sqrt{a^2+r^2}}
  ,\nonumber\\
  \sS{}_{{1}} {}^{0 2} &=& -\sS{}_{{1}} {}^{2 0} =\frac{ a^2 \sqrt{2M r} \sin\theta\cos\theta}{\rho^4\sqrt{a^2+r^2}}
   ,\nonumber\\
 \sS{}_{{1}} {}^{1 3} &=& -\sS{}_{{1}} {}^{3 1} = \frac{ a r \sqrt{2M r}} {\rho^4\sqrt{a^2+r^2}}
    ,\nonumber\\
\sS{}_{{1}} {}^{2 3} &=& -\sS{}_{{1}} {}^{3 2} =  \frac{ a \sqrt{2M r}\cot\theta} {\rho^4\sqrt{a^2+r^2}}
    ,\nonumber\\
\sS{}_{{2}} {}^{0 1} &=& -\sS{}_{{2}} {}^{1 0} =  \frac{ a^2 \sqrt{2M r}\sqrt{a^2+r^2}\sin\theta\cos\theta} {\rho^4}
    ,\nonumber\\
 \sS{}_{{2}} {}^{0 2}& =& -\sS{}_{{2}} {}^{2 0} = -  \frac{ \sqrt{2M r}\l[\rho^2\l(a^2 +3r^2\r)-2r^2\l(a^2+r^2 \r)  \r] } {2r\rho^4\sqrt{a^2+r^2}}
    ,\nonumber\\
\sS{}_{{2}} {}^{1 2} &=& -\sS{}_{{2}} {}^{2 1} = \frac{ M}{a^2 \rho^2}
,\nonumber\\
 \sS{}_{{2}} {}^{1 3} &=& -\sS{}_{{2}} {}^{3 1} =   - \frac{ a \sqrt{2M r}\sqrt{a^2+r^2}\cot\theta} {\rho^4}
 ,\nonumber\\
 \sS{}_{{2}} {}^{2 3} &=& - \sS{}_{{2}} {}^{3 2} =  \frac{ a\sqrt{2M r}\l(\rho^2-2r^2\r) } {2r\rho^4\sqrt{a^2+r^2}}
 ,\nonumber\\
 \sS{}_{{3}} {}^{0 1} &=&  - \sS{}_{{3}} {}^{1 0} = -  \frac{ {2aM }\l[\rho^2\l(a^2 -r^2\r)-2r^2\l(a^2+r^2 \r)  \r] \sin\theta\cos\theta} {\rho^6}
  ,\nonumber\\
 \sS{}_{{3}} {}^{0 2} &=& -\sS{}_{{3}} {}^{2 0} =  -  \frac{ {2Mra^3 } \sin^3\theta\cos\theta} {\rho^6}
   ,\nonumber\\
\sS{}_{{3}} {}^{0 3} &=& - \sS{}_{{3}} {}^{3 0} = - \frac{ \sqrt{2M r}\sqrt{a^2+r^2}} {r\rho^2}
 ,\nonumber\\
\sS{}_{{3}} {}^{1 2} &=& - \sS{}_{{3}} {}^{2 1} = \frac{ a\sqrt{2M r}\sqrt{a^2+r^2}\sin\theta\cos\theta} {\rho^4}
  ,\nonumber\\
  \sS{}_{{3}} {}^{1 3} &=& - \sS{}_{{3}} {}^{3 1} = \frac{ M\l[\rho^2\l(a^2 +3r^2\r)-2r^2\l(a^2+r^2 \r)  \r] } {\rho^6}
    ,\nonumber\\
    \sS{}_{{3}} {}^{2 3} &=& -\sS{}_{{3}} {}^{3 2} =   \frac{ {2Mra^2 } \sin\theta\cos\theta} {\rho^6}\,.
\label{Doran_S}
    \eea

\bibliography{references}

\begin{thebibliography}{10}

\bibitem{BeltranJimenez:2019tjy}
J.~B. Jiménez, L.~Heisenberg, and T.~S. Koivisto.
\newblock {The Geometrical Trinity of Gravity}.
\newblock {\em Universe}, 5(7):173, 2019.

\bibitem{Jimenez:2019ghw}
J.~B. Jim\'enez, L.~Heisenberg, D.~Iosifidis, A.~Jim\'enez-Cano, and T.~S. Koivisto.
\newblock {General teleparallel quadratic gravity}.
\newblock {\em Phys. Lett. B}, 805:135422, 2020.

\bibitem{Bahamonde:2021gfp}
S.~Bahamonde, K.~F. Dialektopoulos, C.~Escamilla-Rivera, G.~Farrugia, V.~Gakis, M.~Hendry, M.~Hohmann, Levi~S. J., J.~Mifsud, and E.~Di~Valentino.
\newblock {Teleparallel gravity: from theory to cosmology}.
\newblock {\em Rept. Prog. Phys.}, 86(2):026901, 2023.

\bibitem{Aldrovandi_Pereira_2013}
R.~Aldrovandi and J.~G. Pereira.
\newblock {\em {Teleparallel Gravity: An Introduction}}.
\newblock Springer, Dordrechts, 2012.

\bibitem{Maluf}
J.~W. Maluf.
\newblock {The teleparallel equivalent of general relativity}.
\newblock {\em Annalen Phys.}, 525:339--357, 2013.

\bibitem{REV_2018}
M.~Kr\v{s}\v{s}\'ak, R.~J. van~den Hoogen, J.~G. Pereira, C.~G. B\"ohmer, and A.~A. Coley.
\newblock {Teleparallel theories of gravity: illuminating a fully invariant approach}.
\newblock {\em Class. Quant. Grav.}, 36(18):183001, 2019.

\bibitem{Ferraro:2006jd}
R.~Ferraro and F.~Fiorini.
\newblock {Modified teleparallel gravity: Inflation without inflaton}.
\newblock {\em Phys. Rev. D}, 75:084031, 2007.

\bibitem{Bengochea:2008gz}
G.~R. Bengochea and R.~Ferraro.
\newblock {Dark torsion as the cosmic speed-up}.
\newblock {\em Phys. Rev. D}, 79:124019, 2009.

\bibitem{Ferraro:2008ey}
R.~Ferraro and F.~Fiorini.
\newblock {On Born-Infeld Gravity in Weitzenbock spacetime}.
\newblock {\em Phys. Rev. D}, 78:124019, 2008.

\bibitem{Linder:2010py}
E.~V. Linder.
\newblock {Einstein's Other Gravity and the Acceleration of the Universe}.
\newblock {\em Phys. Rev. D}, 81:127301, 2010.
\newblock [Erratum: Phys. Rev. D 82,109902(2010)].

\bibitem{Cai:2015emx}
Y.-F. Cai, S.~Capozziello, M.~De~Laurentis, and E.~N. Saridakis.
\newblock {f(T) teleparallel gravity and cosmology}.
\newblock {\em Rept. Prog. Phys.}, 79(10):106901, 2016.

\bibitem{Hayashi:1979qx}
K.~Hayashi and T.~Shirafuji.
\newblock {New General Relativity}.
\newblock {\em Phys. Rev. D}, 19:3524--3553, 1979.
\newblock [Addendum: Phys.Rev.D 24, 3312--3314 (1982)].

\bibitem{Geng:2011aj}
C.-Q. Geng, C.-C. Lee, E.~N. Saridakis, and Y.-P. Wu.
\newblock {\textquotedblleft{}Teleparallel\textquotedblright{} dark energy}.
\newblock {\em Phys. Lett. B}, 704:384--387, 2011.

\bibitem{Maluf:2011kf}
J.~W. Maluf and F.~F. Faria.
\newblock {Conformally invariant teleparallel theories of gravity}.
\newblock {\em Phys. Rev. D}, 85:027502, 2012.

\bibitem{Bahamonde:2017wwk}
S.~Bahamonde, C.~G. B\"ohmer, and M.~Kr\v{s}\v{s}\'ak.
\newblock {New classes of modified teleparallel gravity models}.
\newblock {\em Phys. Lett. B}, 775:37--43, 2017.

\bibitem{Hohmann:2017duq}
M.~Hohmann, L.~J\"arv, M.~Kr\v{s}\v{s}\'ak, and C.~Pfeifer.
\newblock {Teleparallel theories of gravity as analogue of nonlinear electrodynamics}.
\newblock {\em Phys. Rev. D}, 97(10):104042, 2018.

\bibitem{Heisenberg:2023lru}
L.~Heisenberg.
\newblock {Review on f(Q) gravity}.
\newblock {\em Phys. Rept.}, 1066:1--78, 2024.

\bibitem{Moller1961}
C.~Moller.
\newblock Further remarks on the localization of the energy in the general theory of relativity.
\newblock {\em Annals Phys.}, 12(1):118--133, 1961.

\bibitem{M_2M}
F.~I. Mikhail, M.~I. Wanas, Ahmed Hindawi, and E.~I. Lashin.
\newblock {Energy momentum complex in Moller's tetrad theory of gravitation}.
\newblock {\em Int. J. Theor. Phys.}, 32:1627--1642, 1993.

\bibitem{Maluf0704}
J.~W. Maluf, F.~F. Faria, and S.~C. Ulhoa.
\newblock {On reference frames in spacetime and gravitational energy in freely falling frames}.
\newblock {\em Class. Quant. Grav.}, 24:2743--2754, 2007.

\bibitem{9}
J.~W. Maluf, M.~V.~O. Veiga, and J.~F. da~Rocha-Neto.
\newblock {Regularized expression for the gravitational energy-momentum in teleparallel gravity and the principle of equivalence}.
\newblock {\em Gen. Rel. Grav.}, 39:227--240, 2007.

\bibitem{Maluf:2018coz}
J.~W. Maluf, S.~C. Ulhoa, J.~F. da~Rocha-Neto, and F.~L. Carneiro.
\newblock {Difficulties of Teleparallel Theories of Gravity with Local Lorentz Symmetry}.
\newblock {\em Class. Quant. Grav.}, 37(6):067003, 2020.

\bibitem{Capozziello2018}
S.~Capozziello, M.~Capriolo, and M.~Transirico.
\newblock The gravitation energy–momentum pseudotensor: The cases of f(r) and f(t) gravity.
\newblock {\em International Journal of Geometric Methods in Modern Physics.}, 15(1):1850164, 2018.

\bibitem{Krssak}
M.~Kr\v{s}\v{s}\'ak.
\newblock On the problem of energy-momentum in teleparallel gravity.
\newblock {\em Conference: Teleparallel Universes in Salamanca (Salamanca, 26–28 November 2018).}

\bibitem{Obukhov+}
T.~G. Lucas, Y.~N. Obukhov, and J.~G. Pereira.
\newblock {Regularizing role of teleparallelism}.
\newblock {\em Phys. Rev. D}, 80:064043, 2009.

\bibitem{Obukhov_2006}
Y.~N. Obukhov and G.~F. Rubilar.
\newblock {Covariance properties and regularization of conserved currents in tetrad gravity}.
\newblock {\em Phys. Rev. D}, 73:124017, 2006.

\bibitem{Obukhov_Rubilar_Pereira_2006}
Y.~N. Obukhov, G.~F. Rubilar, and J.~G. Pereira.
\newblock {Conserved currents in gravitational models with quasi-invariant Lagrangians: Application to teleparallel gravity}.
\newblock {\em Phys. Rev. D}, 74:104007, 2006.

\bibitem{EPT19}
E.~D. Emtsova, A.~N. Petrov, and A.~V. Toporensky.
\newblock {Conserved currents and superpotentials in teleparallel equivalent of GR}.
\newblock {\em Class. Quant. Grav.}, 37(9):095006, 2020.

\bibitem{EPT_2020}
E.~D. Emtsova, A.~N. Petrov, and A.~V. Toporensky.
\newblock {On conservation laws in teleparallel gravity}.
\newblock {\em J. Phys. Conf. Ser.}, 1557(1):012017, 2020.

\bibitem{EP:2021snt}
E.~D. Emtsova and A.~N. Petrov.
\newblock {A moving black hole in TEGR as a moving matter ball}.
\newblock {\em Space, Time and Fundamantal Interactions.}, (39):18--25, 2022.

\bibitem{EP:2022ohe}
E.~D. Emtsova and A.~N. Petrov.
\newblock {On gauges for a moving black hole in TEGR}.
\newblock {\em Gen. Rel. Grav.}, 54(10):114, 2022.

\bibitem{EKPT_2021}
E.~D. Emtsova, M.~Kr\v{s}\v{s}\'ak, A.~N. Petrov, and A.~V. Toporensky.
\newblock {On conserved quantities for the Schwarzschild black hole in teleparallel gravity}.
\newblock {\em Eur. Phys. J. C}, 81(8):743, 2021.

\bibitem{EKPT_2021a}
E.~D. Emtsova, M.~Kr\v{s}\v{s}\'ak, A.~N. Petrov, and A.~V. Toporensky.
\newblock {On the Schwarzschild solution in TEGR}.
\newblock {\em J. Phys. Conf. Ser.}, 2081(1):012017, 2021.

\bibitem{EPT:2022uij}
E.~D. Emtsova, A.~N. Petrov, and A.~V. Toporensky.
\newblock {Conserved quantities in STEGR and applications}.
\newblock {\em Eur. Phys. J. C}, 83(5):366, 2023.

\bibitem{EPT:2023hbc}
E.~D. Emtsova, A.~N. Petrov, and A.~V. Toporensky.
\newblock {The equivalence principle for a plane gravitational wave in torsion-based and non-metricity-based teleparallel equivalents of general relativity}.
\newblock {\em Eur. Phys. J. C}, 84(3):215, 2024.

\bibitem{Nester_2020}
M.~Blagojevi\'c and J.~M. Nester.
\newblock {Local symmetries and physical degrees of freedom in gravity: A Dirac-Hamiltonian constraint analysis}.
\newblock {\em Phys. Rev. D}, 102:064025, 2020.

\bibitem{Golovnev_2024}
A.~Golovnev.
\newblock {Degrees of Freedom in modified Teleparallel Gravity}.
\newblock {\em Ukrainian J. Phys.}, 69:456, 2024.

\bibitem{Ambrosio_2023}
F.~D'Ambrosio, L.~Heisenberg, and S.~Zentarra.
\newblock {Hamiltonian Analysis of $f(Q)$ Gravity and the Failure of the Dirac-Bergmann Algorithm for Teleparallel Theories of Gravity}.
\newblock {\em arXiv:2308.02250 [gr-qc]}.

\bibitem{Capozziello:2022zzh}
S.~Capozziello, V.~De~Falco, and C.~Ferrara.
\newblock {Comparing equivalent gravities: common features and differences}.
\newblock {\em Eur. Phys. J. C}, 82(10):865, 2022.

\bibitem{Xu:2006ki}
S.-X. Xu and J.-L. Jing.
\newblock {Energy of general 4-dimensional stationary axisymmetric spacetime in the teleparallel geometry}.
\newblock {\em Class. Quant. Grav.}, 23:4659--4672, 2006.

\bibitem{daRocha-Neto:2002cvu}
J.~F. da~Rocha-Neto and K.~H. Castello-Branco.
\newblock {Gravitational energy of Kerr and Kerr anti-de Sitter space-times in the teleparallel geometry}.
\newblock {\em JHEP}, 11:002, 2003.

\bibitem{Maluf:2002zc}
J.~W. Maluf, J.~F. da~Rocha-Neto, T.~M.~L. Toribio, and K.~H. Castello-Branco.
\newblock {Energy and angular momentum of the gravitational field in the teleparallel geometry}.
\newblock {\em Phys. Rev. D}, 65:124001, 2002.

\bibitem{Maluf:2000cc}
J.~W. Maluf, J.~F. da~Rocha-Neto, T.~M.~L. Toribio, and K.~H. Castello-Branco.
\newblock {Energy momentum of the gravitational field in the teleparallel geometry}.
\newblock In {\em {9th Marcel Grossmann Meeting on Recent Developments in Theoretical and Experimental General Relativity, Gravitation and Relativistic Field Theories (MG 9)}}, pages 1043--1044, 7 2000.

\bibitem{Maluf:1996kx}
J.~W. Maluf, E.~F. Martins, and A.~Kneip.
\newblock {Gravitational energy of rotating black holes}.
\newblock {\em J. Math. Phys.}, 37:6302--6310, 1996.

\bibitem{Maluf:1995rd}
J.~W. Maluf and A.~Kneip.
\newblock {Localization of energy for Kerr black hole}.
\newblock {\em Arxiv:gr-qc/9504011}.

\bibitem{Nashed:2006yw}
G.~G.~L. Nashed.
\newblock {Kerr-Newman Solution and Energy in Teleparallel Equivalent of Einstein Theory}.
\newblock {\em Mod. Phys. Lett. A}, 22:1047--1056, 2007.

\bibitem{Bejarano2015}
C.~Bejarano, R.~Ferraro, and M.~J. Guzmán.
\newblock Kerr geometry in $f(t)$ gravity.
\newblock {\em Eur. Phys. J. C}, 75:77, 2015.

\bibitem{Gomes:2022vrc}
D.~A. Gomes, J.~J. Jim\'enez, and T.~S. Koivisto.
\newblock {Energy and entropy in the Geometrical Trinity of gravity}.
\newblock {\em Arxiv:2205.09716, [gr-qc]}.

\bibitem{BeltranJimenez:2019tme}
B.~J. Jim\'enez, L.~Heisenberg, T.~S. Koivisto, and S.~Pekar.
\newblock {Cosmology in $f(Q)$ geometry}.
\newblock {\em Phys. Rev. D}, 101(10):103507, 2020.

\bibitem{Rodrigues:2013ifa}
M.~E. Rodrigues, M.~J.~S. Houndjo, J.~Tossa, D.~Momeni, and R.~Myrzakulov.
\newblock {Charged Black Holes in Generalized Teleparallel Gravity}.
\newblock {\em JCAP}, 11:024, 2013.

\bibitem{Doran:1999gb}
C.~Doran.
\newblock {A New form of the Kerr solution}.
\newblock {\em Phys. Rev. D}, 61:067503, 2000.

\bibitem{Landau_Lifshitz_1975}
L.~D. Landau and E.~M. Lifschits.
\newblock {\em {The Classical Theory of Fields}}.
\newblock Pergamon Press, 1975.

\bibitem{Golovnev:2017dox}
A.~Golovnev, T.~Koivisto, and M.~Sandstad.
\newblock {On the covariance of teleparallel gravity theories}.
\newblock {\em Class. Quant. Grav.}, 34(14):145013, 2017.

\bibitem{Mitskevich_1969}
N.~V. Mitskevich.
\newblock {\em {Physical Fields in General Relativity Theory}}.
\newblock Nauka, Moscow, 1969.

\bibitem{Petrov_KLT_2017}
A.~N. Petrov, S.~M. Kopeikin, R.~R. Lompay, and B.~Tekin.
\newblock {\em {Metric Theories of Gravity: Perturbations and Conservation Laws}}, volume~38 of {\em De Gruyter Studies in Mathematical Physics}.
\newblock De Gruyter, 4 2017.

\bibitem{Adak:2011ltj}
M.~Adak and C.~Pala.
\newblock {A novel approach to autoparallels for the theories of symmetric teleparallel gravity}.
\newblock {\em J. Phys. Conf. Ser.}, 2191(1):012017, 2022.

\bibitem{BeltranJimenez:2022azb}
J.~B. Jiménez and T.~S. Koivisto.
\newblock Lost in translation: The abelian affine connection (in the coincident gauge).
\newblock {\em International Journal of Geometric Methods in Modern Physics.}, 19(7):2250108, 2022.

\bibitem{KBL_1997}
J.~Katz, J.~Bi\v{c}\'ak, and D.~Lynden-Bell.
\newblock Relativistic conservation laws and integral constraints for large cosmological perturbations.
\newblock {\em Phys. Rev. D}, 55:5957, 1997.

\bibitem{Katz1985}
J.~Katz.
\newblock A note on komar's anomalous factor.
\newblock {\em Class. Quantum. Grav.}, 2(3):423, 1985.

\bibitem{Painlev}
P.~Painlevé.
\newblock {La mécanique classique et la théorie de la relativité}.
\newblock {\em C. R. Acad. Sci. (Paris)}, 173:677–680, 1921.

\bibitem{Gullstrand}
A.~Gullstrand.
\newblock {Allgemeine Lösung des statischen Einkörperproblems in der Einsteinschen Gravitationstheorie}.
\newblock {\em Arkiv för Matematik, Astronomi och Fysik}, 16 (8):1–15, 1922.

\end{thebibliography}
\bibliographystyle{Style}

\end{document}